\documentclass[11pt]{article}
\pdfoutput=1
\usepackage{amsmath, amsfonts, amssymb}
\usepackage{comment}
\usepackage{graphicx}
\usepackage{psfrag}
\usepackage{amsthm}
\usepackage[usenames,dvipsnames,svgnames,table]{xcolor}
\usepackage{enumerate}
\usepackage{tcolorbox}

\usepackage{soul}
\usepackage{subfig}
\usepackage{mathrsfs}  
\usepackage{a4wide}
\usepackage{booktabs}
\usepackage{tikz}
\usepackage{tikz-cd}
\usepackage{makecell}
\usepackage{color}
\definecolor{dark-gray}{gray}{0.20}
\definecolor{gray}{gray}{0.30}
\definecolor{light-gray}{gray}{0.80}
\definecolor{dark-red}{rgb}{0.7,0,0}
\definecolor{dark-green}{rgb}{0.1,0.4,0}
\definecolor{dark-blue}{rgb}{0.3,0.3,0.7}
\definecolor{light-blue}{rgb}{0.8,0.8,1}
\definecolor{swamp}{RGB}{240, 199, 197}

\usepackage{pifont}
\usepackage{setspace}

\newcommand{\be}{\begin{equation}}
\newcommand{\ee}{\end{equation}}

\def\be{\begin{equation}}
\def\ee{\end{equation}}
\def\bea{\begin{eqnarray}}
\def\eea{\end{eqnarray}}

\newcommand{\dd}{\mathrm{d}}
\newcommand{\K}{Kahler }

\def\simleq{\; \raise0.3ex\hbox{$<$\kern-0.75em
		\raise-1.1ex\hbox{$\sim$}}\; }
\def\simgeq{\; \raise0.3ex\hbox{$>$\kern-0.75em
		\raise-1.1ex\hbox{$\sim$}}\; }

\numberwithin{equation}{section}

\usepackage{jheppub} %jheppub already loads hyperref; load it again and it causes trouble
\hypersetup{
	colorlinks=true,
	linkcolor=dark-blue,
	citecolor=dark-red,
	urlcolor=dark-green,
	linktoc=page,
	pageanchor=false
}
\title{\centering Loops, Local Corrections and Warping\\ 
in the LVS and other Type IIB Models
}

\author{Xin Gao$^1$,}
\author{Arthur Hebecker$^2$,}
\author{Simon Schreyer$^2$,}
\author{ and Gerben Venken$^2$}

\affiliation{$^1$ College of Physics, Sichuan University, Chengdu, 610065, China} 
\affiliation{$^2$ Institute for Theoretical Physics, Heidelberg University,\\
	Philosophenweg 19, 69120 Heidelberg, Germany} 
	
\emailAdd{xingao@scu.edu.cn}
\emailAdd{a.hebecker@thphys.uni-heidelberg.de}
\emailAdd{s.schreyer@thphys.uni-heidelberg.de}
\emailAdd{g.venken@thphys.uni-heidelberg.de}

\abstract{To establish metastable de Sitter vacua or even just scale-separated AdS, control over perturbative corrections to the string-derived leading-order 4d lagrangian is crucial. Such corrections can be classified in three types: First, there are genuine loop effects, insensitive to the UV completion of the 10d theory. Second, there are local $\alpha'$ corrections or, equivalently, 10d higher-dimension operators which may or may not be related to loop-effects. Third, warping corrections affect the 4d Kahler potential but are expected not to violate the 4d no-scale structure. With this classification in mind, we attempt to derive the Berg-Haack-Pajer conjecture for Kahler corrections in type-IIB Calabi-Yau orientifolds and extend it to include further terms.
This is crucial since the interesting applications of this conjecture are in the context of generic Calabi-Yau geometries rather than in the torus-based models from which the main motivation originally stems. As an important by-product, we resolve a known apparent inconsistency between the parametric behaviour of string loop results and field-theoretic expectations. Our findings lead to some interesting new statements concerning loop effects associated with blowup-cycles, loop corrections in fibre inflation, and possible logarithmic effects in the Kahler and scalar potential.
}

\begin{document}

\makeatletter
\let\old@fpheader\@fpheader

\makeatother

\maketitle

%\tableofcontents

%%%%%%%%%%%%%%%%%%%%%%%%%%%%%
%%%%%%%%%%%%%%%%%%%%%%%%%%%%%
\section{Introduction}\label{intro}
%%%%%%%%%%%%%%%%%%%%%%%%%%%%%
%%%%%%%%%%%%%%%%%%%%%%%%%%%%%

The leading paradigm in the search for realistic vacua in the string theory landscape is to start with type-IIB Calabi-Yau orientifold models with O3/O7-planes, to stabilize complex structure moduli by 3-form flux, and only then to deal with the classically flat Kahler moduli space \cite{Dasgupta:1999ss,Giddings:2001yu}. Those flat directions may then be stabilized by nonperturbative effects alone \cite{Kachru:2003aw} or in combination with $\alpha'$ corrections and loop effects \cite{Balasubramanian:2005zx,Conlon:2005ki, Westphal:2006tn}.
In any case, be it as a central ingredient or as a potentially dangerous, subleading effect, perturbative corrections are important in string phenomenology. They affect the scalar potential and hence the prospects of uplifting an initial AdS vacuum to de Sitter -- a key step which is still under debate\footnote{
This 
discussion has recently gained momentum following \cite{Danielsson:2018ztv, Obied:2018sgi}. For some of the latest additions, see e.g.~\cite{Moritz:2017xto,Bena:2018fqc,Carta:2019rhx,Blumenhagen:2019qcg,Kachru:2019dvo, Gao:2020xqh,Demirtas:2021nlu,Bena:2020xrh,Hamada:2021ryq,Junghans:2022exo,Gao:2022fdi}. An important part of the debate is the issue of scale separation \cite{Gautason:2018gln, Lust:2019zwm}.
}. 
Clearly, loop and other perturbative effects also impact models of inflation which use Kahler moduli \cite{Conlon:2005jm, Cicoli:2008gp}. 

Motivated by this situation, we devote the present paper to the study of loop corrections to the type-IIB Kahler moduli Kahler potential in the Calabi-Yau context \cite{vonGersdorff:2005bf, Berg:2005ja, Berg:2005yu, Berg:2007wt,Cicoli:2007xp,Cicoli:2008va,Berg:2014ama,Haack:2018ufg}. The present level of understanding is not satisfactory: While field-theoretic arguments allow one to make a proposal for such corrections in the simplest Calabi-Yau settings \cite{vonGersdorff:2005bf,Cicoli:2007xp}, explicit string loop calculations are available only in torus orbifold geometries \cite{Berg:2005ja,Berg:2005yu}. It has been conjectured how to generalize the latter to Calabi-Yau models \cite{Berg:2007wt,Cicoli:2007xp}, but no derivation for the proposed structure is available. Moreover, there is a seeming inconsistency \cite{Cicoli:2007xp} between field-theoretic and string loop results, which we will resolve in this paper.

In our analysis, we will have the Large Volume Scenario (LVS) \cite{Balasubramanian:2005zx,Conlon:2005ki} in the back of our minds as this is a prototypical example of a model with fluxes where a better understanding of loop and $\alpha'$ corrections is crucial -- see \cite{Conlon:2010ji,Ciupke:2015msa,Minasian:2015bxa, Antoniadis:2018hqy, Weissenbacher:2019mef,Antoniadis:2019rkh,Grimm:2013gma,Grimm:2013bha,Junghans:2014zla,Burgess:2020qsc,Cicoli:2021rub, Burgess:2022nbx, Leontaris:2022rzj} for recent work on loop and $\alpha'$ corrections in this and related settings. However, our findings are not restricted to the LVS and should be relevant more generally in the type-IIB context. 

To explain our approach at a more technical level, let us start by stating the Berg-Haack-Pajer (BHP) conjecture and then describe how, according to our findings, it relates to the three basic types of loop corrections between which we will distinguish. The BHP conjecture proposes two kinds of corrections to the \K potential, scaling like 
\begin{equation}
    \delta K^{KK}_{(g_s)} \sim \sum\limits_{a}\frac{g_s \mathcal{T}_a(t^i)}{\mathcal{V}}\,, \quad\quad\quad \delta K^{W}_{(g_s)} \sim \sum\limits_{a}\frac{1 }{\mathcal{I}_a(t^i)\mathcal{V}}\,.
    \label{bhp}
\end{equation}
Here $\mathcal{T}_a$ and $\mathcal{I}_a$ are linear combinations of 2-cycle \K moduli $t^i$, the volume is ${\cal V}=\kappa_{ijk} t^i t^j t^k/6$, and we recall that the proper Kahler variables are the complex 4-cycle moduli, with real parts $\tau_i=\partial {\cal V}/\partial t^i$. While, as we will see momentarily, our results in part deviate from \eqref{bhp}, it is nevertheless a good starting point for organizing our discussion.

Next, we clarify the origin and fix terminology for three different kinds of loop corrections:

First, there are {\bf genuine loop corrections} which arise from integrating out the tower of KK modes (4d perspective) or from loops of 10d or brane-localized fields propagating in the compact space (10d perspective). Their distinguishing feature is their non-locality in the higher-dimensional theory: They can not be associated with local operators in 10d or on a brane. In this sense, they are analogous to the Casimir energy\footnote{As a result, our analysis may be relevant for compactification schemes directly relying on Casimir energy, see e.g.~\cite{DeLuca:2021pej}.}, which arises in geometries with two separated surfaces but can {\it not} be encoded in a local operator on either surface or in the space between them. 

The genuine loop corrections may be thought of as coming from the interacting 4d field theory of moduli and KK modes. In this theory, 3-vertices are suppressed by $1/M_4$. 
Accordingly, genuine 1-loop effects correct the Kahler moduli kinetic terms as
\begin{equation}
    \left(1+\frac{M_{KK}^2}{M_4^2}\right)\frac{1}{\tau^2}\partial_\mu\tau\partial^\mu\tau
     \qquad \mbox{or, more generally,}\qquad
    \left(1+\frac{M_{KK}^2}{M_4^2}\right)K_{ij}\partial_\mu\tau^i\partial^\mu\tau^j
    \,.
    \label{genuinescaling}
\end{equation}
Here the factor $M_{KK}^2$ appears on dimensional grounds since, similarly to Casimir energy calculations, no UV mass scales are involved.

It is easy to see that $M_{KK}^2/M_4^2$ is a homogeneous function of degree -2 in Einstein-frame 4-cycle volumes. Equivalently, we can say that the correction appears at order $\alpha'^4g_s^2$, such that its scaling agrees with that of the second term of the BHP conjecture \eqref{bhp}. Our previously mentioned inconsistency is then the apparent absence of the first term in \eqref{bhp} in the field-theoretic approach. Moreover, we will argue that the functions $\mathcal{I}_a$ in the Calabi-Yau case are not necessarily linear in 2-cycles. Instead, the additional dependence on ratios of cycles is expected.
In Sect.~\ref{loops}, we discuss the genuine loop corrections in detail and also derive \eqref{genuinescaling} using Feynman diagrams.

Second, there are {\bf local $\alpha'$ corrections} or, in more precise language, corrections coming from higher-dimension local operators in 10d, on branes and O-planes, or on their intersection loci.
We use the adjective `local' to distinguish them from other effects, such as genuine loop corrections, which also induce 4d EFT operators suppressed by $\alpha'$. It is important to note that local $\alpha'$ corrections may receive contributions from the high-momentum region of loop integrals. There is in particular no clean separation between local $\alpha'$ corrections which are part of the classical action and the counterterms needed to renormalize the loops. It appears natural to us to collect all corrections which can be associated with higher-dimension local operators, be they fundamental or loop-induced, under the name `local $\alpha'$ corrections'.

Local $\alpha'$ corrections appear at different order in $\alpha'$ since the underlying higher-dimension operators come with different $\alpha'$ suppression factors.
Crucially, such local corrections at order $\alpha'^2$ can explain the first term in \eqref{bhp} and thus resolve the above puzzle. Other local $\alpha'$ corrections contribute to the second term in \eqref{bhp}, with or without additional $g_s$ factors. This depends on whether the operator in question appears at the string tree level or at higher-loop order.

An important result of our paper, for which we argue in Sect.~\ref{dbranes}, is the general expectation that marginal local operators (appearing at order $\alpha'^4$) introduce logarithmic corrections to the \K potential. Examples for this would be, if existent, the $R_8^4$ operator on a D7-brane/O7-plane and the $R_6^3$ operator on the intersection locus between D7-branes/O7-planes. In Sect.~\ref{membranesection}, we will deal more generally with loop corrections induced by localized objects and extend the results to multiple Kahler moduli.

Finally, there are {\bf warping corrections} or, more generally, corrections due to the classical backreaction of the background geometry. These can not be cleanly separated from string loop effects since, in the regime where the worldsheet is a long cylinder, the string loop encodes the effects of light 10d fields propagating at tree-level. In our 10d EFT approach, such corrections have to be viewed as classical rather than loop-induced.

As is well known (and reviewed in Sect.~\ref{warping}) warping corrects the Kahler potential by a series of terms $1/\tau^n$, starting at $n=1$. The complete series does not affect the scalar potential since warping respects the no-scale structure \cite{Giddings:2001yu}. The $n=2$ term matches parametrically the second term in \eqref{bhp}.

In Sect.~\ref{applications} we work out the explicit form of loop corrections for a blowup modulus (see also \cite{Roth}) and for fibred geometries. Before concluding in Sect.~\ref{discussion}, we devote Sect.~\ref{generalissues} to some further applications where loop corrections can be important: The parametric control of the LVS \cite{Junghans:2022exo, Gao:2022fdi}, the control of KKLT with many moduli and small 2- or 4-cycles \cite{Gao:2020xqh, Demirtas:2021nlu}, and the possible presence of dominant log-corrections to the Kahler potential \cite{Antoniadis:2018hqy, Weissenbacher:2019mef, Antoniadis:2019rkh, Burgess:2022nbx, Leontaris:2022rzj}.
Appendix \ref{martucci} contains more details concerning our discussion of warping corrections in Sect.~\ref{warping}, following mainly \cite{Martucci:2014ska}.

\begin{table}[!h]\centering
	\small
	\caption{Some of the key corrections discussed in the paper and their effect on the \K and scalar potential. The functions $f_{-\lambda}$, $h_{-\lambda}$ are homogeneous of degree $-\lambda$ in 4-cycles and $L$ is the typical length scale of the internal manifold.}
	
	\vspace{.3cm}
	\label{tab:correction_summary}
		\begin{tabular}{c|c|c|c|c}
			Correction type & \makecell{Discussed\\in Section} & Induced by &  \makecell{Correction to\\\K potential} & \makecell{Correction to\\scalar potential}\\
			\hline
			\midrule
			Genuine loops \cite{vonGersdorff:2005bf}&\ref{loops} and \ref{generalization} & - & $f_{-2}$ & $|W_0|^2 g_s\times h_{-5} $\\
			\midrule
			BBHL+1-loop \cite{Antoniadis:1997eg,Becker:2002nn}& \ref{localalphacorrections}& $\frac{M_{10}^2}{g_s^{3/2}}(1+g_s^2)R^4_{10}$&\makecell{$ (g_s^{-1/2}+g_s^{3/2})$\\ $\times f_{-3/2}$} &\makecell{$|W_0|^2(g_s^{-3/2} + g_s^{1/2})$ \\$ \times h_{-9/2} $}\\
			\midrule
			\makecell{Non-intersecting \\ D7/O7 (partly) \cite{Grimm:2013gma,Grimm:2013bha}}& \ref{dbranes}&$M_{10}^4(1+g_s)R_8^2$ &$(0+g_s)\times f_{-1} $&$|W_0|^2 g_s^3\times h_{-5} $\\
			\midrule
			\makecell{Log-Correction \\ on D7/O7}& \ref{dbranes}& $R^4_{8}$ & \makecell{ $\ln(M_{10}g_s^{1/4}L)$\\ $\times f_{-2}$} &\makecell{ $|W_0|^2 g_s \ln(M_{10}g_s^{1/4}L)$\\$\times h_{-5} $}\\
			\midrule
			\makecell{Intersecting D7/O7 \\ \cite{Grimm:2013gma,Grimm:2013bha,Junghans:2014zla,Epple:2004ra,Haack:2015pbv}}&\ref{intersectingbranes} &$M_{10}^4(1+g_s)R_6$ & $(0+g_s)\times f_{-1}$ &$|W_0|^2 g_s^3\times h_{-5} $\\
			\midrule
			\makecell{Log-Correction \\ on intersecting D7/O7}& \ref{intersectingbranes}& $R^3_{6}$ & \makecell{ $\ln(M_{10}g_s^{1/4}L)$\\ $\times f_{-2}$} &\makecell{ $|W_0|^2 g_s \ln(M_{10}g_s^{1/4}L)$\\$\times h_{-5} $}\\
			\bottomrule
	\end{tabular}
\end{table}
\newpage
Table \ref{tab:correction_summary} provides a partial list of the genuine loop and local $\alpha'$ corrections considered in this paper\footnote{We do not include warping corrections since they do not affect the scalar potential.}. In particular, concerning the operators on branes and their intersections, we display only the lowest-dimension and marginal operators. It is convenient to express the correction to the \K and scalar potential in terms of homogeneous functions of a certain degree in the \K moduli since the detailed dependence on ratios of 4-cycle volumes is known only in special cases.
We emphasize in particular the corrections induced by an $R_8^4$ term, potentially present on D7-branes and O7-planes, which has to our knowledge not been considered before. Being marginal and hence probably log-divergent, this operator induces a correction to the scalar potential which is leading compared to the  corrections following from the BHP conjecture. It is therefore important for cosmological applications like Fibre Inflation (to be discussed in Sect.~\ref{fibreinflation}) or moduli stabilization scenarios involving loop effects.

%%%%%%%%%%%%%%%%%%%%%%%%%%%%%
%%%%%%%%%%%%%%%%%%%%%%%%%%%%%
\section{Basics of Loop Corrections - The Single Modulus Case}\label{loops}
%%%%%%%%%%%%%%%%%%%%%%%%%%%%%
%%%%%%%%%%%%%%%%%%%%%%%%%%%%%

%%%%%%%%%%%%%%%%%%%%%%%%%%%%%
\subsection{Naive Power Counting} \label{dimsingle}
%%%%%%%%%%%%%%%%%%%%%%%%%%%%%

Our goal is a better understanding of the role of loop corrections in type-IIB. Since exact string loop calculations for Calabi-Yau manifolds are not feasible, we will try to develop the parametric estimates based on dimensional analysis as suggested in \cite{vonGersdorff:2005bf}. We will later on compare our findings with the exact torus-orbifold results of \cite{Berg:2005ja} and the corresponding Calabi-Yau form of such corrections conjectured in \cite{Berg:2007wt}.

Let us start from the bosonic part of the Einstein-frame type-IIB action (see e.g.~\cite{Polchinski:1998rr}),
\begin{equation}
S_{\text{EF}}=  \frac{1}{2\kappa_{10}^2}\int\dd^{10}x \sqrt{-g} \left[ R_{10} -\frac{\partial_M\tau\partial^M\overline{\tau}}{2\left(\text{Im}\tau\right)^2} -\frac{G_{(3)}\cdot\overline{G}_{(3)}}{12\text{Im}\tau} - \frac{\tilde{F}_{(5)}^2}{4\cdot5!}\right] + S_{\text{CS}}\,,
\label{2baction}
\end{equation}
where $2\kappa_{10}^2 = (2\pi)^7\alpha'^{4}$, and $S_{\text{CS}}$ the Chern-Simons term. We compactify on a Calabi-Yau orientifold with O3/O7 planes and local tadpole cancellation by D3/D7-branes, without fluxes.\footnote{In the case of D3-branes, this is only possible in a supergravity toy model, where one can place one fourth of a D3 on each O3-plane to cancel its tadpole. In string theory, one can at best place one D3 on every fourth O3. This implies warping corrections, to be discussed below. In the D7/O7 case, the curvature of the brane and O-plane induces a D3 tadpole, such that our analysis without fluxes is, once again, in most cases only an approximation.}
The corresponding metric can be written as
\begin{equation}
\dd s^2=g_{\mu\nu}\dd x^\mu\dd x^\nu + L(x)^2 \tilde{g}_{mn}\dd y^m\dd y^n,
\label{ansatz}
\end{equation}
with $\mu,\nu\in\{0,..,3\}$, $\,\,m,n\in\{4,..,9\}$ and a Calabi-Yau metric $\tilde{g}_{mn}$ normalized such that the compact space has unit volume. The physical Einstein-frame volume is hence given by $\mathcal{V}=L^6$. The resulting action in 4d Jordan-Brans-Dicke frame (Jordan frame for short) reads
\begin{equation}
S_{\text{JBD}} = \frac{1}{2\kappa_{10}^2}\int\dd^4x \sqrt{-g}L^6\left[ R_4+6(6-1) \frac{\left(\partial  L\right)^2}{L^2} +\cdots \right].
\label{S}
\end{equation}
where we only display the Einstein-Hilbert and volume Kahler-modulus kinetic terms.

Postponing a more careful, Feynman-diagram-based derivation to Sect.~\ref{feynman}, we first provide a simple dimensional argument for the parametric behaviour of the genuine loop corrections to \eqref{S}:
At one loop, such corrections come from integrating out the tower of all KK modes. The total UV divergence is absorbed in 10d in a renormalization of $M_{10}^4$. The finite piece knows only about a single dimensionful parameter, the length scale $L$ which governs the KK masses. Hence, on dimensional grounds, the corrections read
\begin{equation}
\Delta S_{\text{JBD}} =\int \dd^4x \sqrt{-g} \left( \frac{b_0}{L^2}R_4+\frac{b_1}{L^4}(\partial L)^2 \right),
\label{sloopcorr}
\end{equation}
where $b_0,b_1$ are $\mathcal{O}(1)$ numerical coefficients  depending on the specific Calabi-Yau and, if present, on complex structure moduli.\footnote{These coefficients can be large if the number of light fields, including in particular complex structure moduli, is large \cite{deAlwis:2021zab}. It has been suggested in \cite{deAlwis:2021zab} to use such large loop corrections to uplift from AdS to de Sitter. We consider it safer to include the loop effects as corrections to the Kahler potential, and then to study the minima of the resulting supergravity scalar potential. The validity of such an approach, even concerning loops with fields below the SUSY breaking scale, has recently been emphasized in \cite{Burgess:2021juk}.
}
Since we perform our analysis of loop corrections to the Kahler potential in a pure Calabi-Yau orientifold, without fluxes, the complex structure moduli are massless. We treat their vevs as parameters and their fluctuations as light fields, running in the loop just like the Kahler modulus, the 4d graviton and the KK modes. We expect flux-induced modifications of the loop corrections to the Kahler potential to be subleading. To see this, recall that the spacing of KK towers is set by $m_{KK}^2\sim 1/L^2$. The lowest level of many of the towers is zero, making the corresponding fields moduli. Three-form fluxes thread 3-cycles and hence scale as $1/L^3$. They induce an energy density $G_3^2\sim 1/L^6$ which depends e.g.~on the 10d metric and hence provides as mass correction for the 10d-metric KK tower. One obtains $\delta m_{KK}^2\sim 1/L^6$, which is clearly subleading. The above is consistent with the well-known fact that complex structure moduli masses scales as $m_{cs}\sim 1/L^3$.

The sum of \eqref{S} and \eqref{sloopcorr} can be translated to 4d Einstein frame. In addition, we trade $L$ for the dimensionless 4-cycle variable $\tau=M_{10}^4L^4/(2\pi)^{4}=L^4/(l_s^4g_s)$, with $l_s=2\pi\sqrt{\alpha'}$. This corresponds to common conventions for measuring 4-cycle volumes in type-IIB. The result is
\begin{equation}
\left(S+\Delta S\right)_\text{EF} = \frac{M_{4}^2}{2}\int \dd^4x \sqrt{-g} \left[ R_4+\left(a_2 \frac{\left(\partial  \tau\right)^2}{\tau^2} +b_2\frac{(\partial \tau)^2}{\tau^4} \right)\right],
\label{efaction}
\end{equation}
with $M_{4}$ the 4d Planck mass, $a_2=-3/2$, and $b_2=(114b_0+b_1)/(32\pi)$ which again depend on complex structure moduli. To make contact with the Kahler potential, we have to interpret \eqref{efaction} as a 4d SUSY action, with $\tau=\mbox{Re}\,T$ and $T$ the complexified Kahler modulus. Thus,
\begin{equation}
({\cal L}+\Delta {\cal L})_\text{EF}=M_{4}^2\left(-\frac{3}{(T+\overline{T})^2} \partial T \cdot \partial \overline{T} + \frac{16b_2}{(T+\overline{T})^4}\partial T \cdot \partial \overline{T}\right),
\label{correctiont}
\end{equation}
where we can identify the prefactors as second derivatives of the Kahler potential $K$ and its loop correction $\delta K$. Hence $\delta K/M_4^2 = 8b_2/3(T+\overline{T})^2$, which induces a term in the scalar potential of order $\mathcal{V}^{-10/3}$. This matches the winding correction of the BHP conjecture \cite{Berg:2007wt}, but it does not capture the leading KK correction. We will resolve this issue in Sect.~\ref{warping} and \ref{comparison}.

%%%%%%%%%%%%%%%%%%%%%%%%%%%%%
\subsection{Support by Feynman-Diagram Calculations}\label{feynman}
%%%%%%%%%%%%%%%%%%%%%%%%%%%%%

We now verify the results of \cite{vonGersdorff:2005bf} reviewed in Sect.~\ref{dimsingle} using a more explicit Feynman diagram argument. We follow the literature on `extra dimensions' \cite{Giudice:1998ck, Han:1998sg, Contino:2001nj}, where Kaluza-Klein (KK) expansions are performed in simple geometries, as well as more general and recent studies \cite{Hinterbichler:2013kwa,Brown:2013mwa,deRham:2014zqa,Braun:2008jp,Ashmore:2020ujw}. Of course, we can not be fully explicit in our Calabi-Yau situation.

We start by expanding the metric as $g_{MN}=g^{(0)}_{MN}+\kappa_{10}h_{MN}$. Here $g^{(0)}_{MN}$ denotes the background metric \eqref{ansatz}, but with $L(x)$ replaced by a constant which, by slight abuse of notation, we call $L$. In other words, we write $L(x)=L+\delta L(x)$ in \eqref{ansatz}, treating the volume modulus $\delta L(x)$ as part of the metric fluctuation $h_{MN}$. In order to obtain a 4d action from which Feynman diagrams can be read off, we KK expand all fields in terms of eigenfunctions of their corresponding Laplace operator\footnote{By this we mean the Laplace-Beltrami operator for scalar fields, the Laplace-de-Rham operator for $p$-forms and the Laplace-Lichnerowicz operator for the graviton (in general for symmetric tensors).} on the Calabi-Yau. In the following we will focus on the 4d graviton and its massive spin-2 modes as an example but similar terms can be written down for all bulk fields in \eqref{2baction} and their KK modes. We first diagonalize the action, eliminating the mixing of graviton modes with scalars and vectors arising from the 10d metric. Focusing on the spin-2 part, the action then reads \cite{Hinterbichler:2013kwa,Brown:2013mwa,Giudice:1998ck,deRham:2014zqa,Braun:2008jp,Ashmore:2020ujw}
	\begin{equation}
		\begin{split}
			S=  \int\dd^4x \Biggl[ \sum\limits_{a} \Biggl( &\frac{1}{2} h^{*\mu\nu,a} \left(\Box+m_{a}^2\right)h^{a}_{\mu\nu}-\frac{1}{2} h^{*\mu,a}_{~~~~\mu} \left(\Box+m_{a}^2\right)h^{\nu,a}_{~~\nu}\\
			& +h^{*\mu\nu,a}\partial_\mu\partial_\nu h^{\lambda,a}_{~~\lambda} - h^{*\mu\nu,a}\partial_\mu\partial_\lambda h^{\lambda,a}_{~~\nu} + c.c.+ ... \Biggr)\\ &+\frac{1}{M_{4}}\sum\limits_{a_1,a_2,a_3} V_3\left[ h^{a_1}_{\mu\nu}, h^{a_2}_{\rho\sigma}, h^{a_3}_{\lambda\alpha}\right] + \frac{1}{M_4^2}V_4[...] + ... \Biggr],
		\end{split}
		\label{kkreduction}
	\end{equation}
where $h^{a}_{\mu\nu}$ are the 4d graviton modes with $a$ labeling the eigenfunctions $\psi_a$ of the Laplace-Beltrami operator with eigenvalues $m_a^2$. In order to obtain \eqref{kkreduction}, one has to use the relation
    \begin{equation}
        h_{\mu\nu} = \sum\limits_{a\neq0} h^a_{\mu\nu} \psi_a + \frac{1}{L^3} h^0_{\mu\nu}
        \label{hexpanded}
    \end{equation}
between the 4d part of the 10d graviton $h_{\mu\nu}$, the 4d graviton $h^0_{\mu\nu}$, and its massive KK modes $h^a_{\mu\nu}$ \cite{Hinterbichler:2013kwa}. To understand the mass dimensions of our fields, recall that our background metric $g_{MN}^{(0)}$ and its correction $\kappa_{10}h_{MN}$ are dimensionless. Hence $[h_{MN}]=-[\kappa_{10}]=4$. Correspondingly, the l.h.~side of \eqref{hexpanded} has mass dimension 4, $[\psi_a]=[1/L^3]=3$, and the fields $h_{\mu\nu}^a$ and $h_{\mu\nu}^0$ have the canonical mass dimension one of 4d bosonic fields. This is consistent with \eqref{kkreduction}.

\begin{figure}[ht]
	\centering
	\includegraphics[width=1\textwidth]{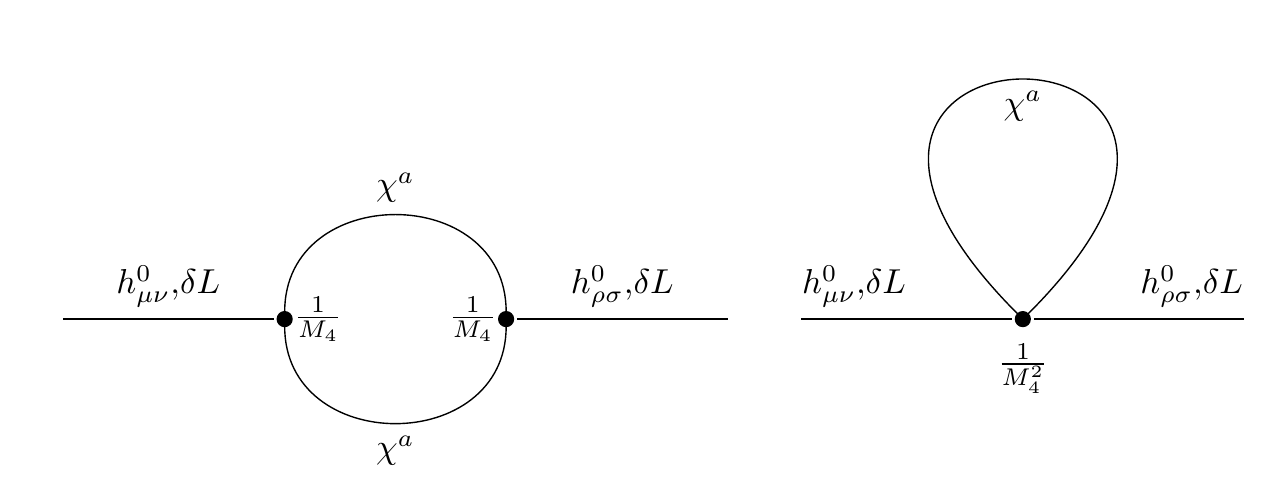}
	\caption{Self-energy diagrams correcting the propagator of the massless 4d graviton and the volume modulus. The 3-vertex is suppressed by $1/M_{4}$ and the 4-vertex by $1/M_{4}^2$. The field $\chi^{a}$ symbolizes all fields with their KK towers, including the massless moduli and ghost fields.}
	\label{loopdiagrams}
\end{figure}

The functionals $V_3$ and $V_4$ are sums of cubic and quartic terms in the $h_{\mu\nu}^{a}$. Each term contains two derivatives. When deducing Feynman rules, $V_3$ and $V_4$ will give 3- and 4-vertices. We have for brevity suppressed the arguments of $V_4$ -- they coincide with those of $V_3$. The ellipsis at the very end of \eqref{kkreduction} stands for higher vertices as well as for analogous quadratic and higher-order action pieces involving all other modes of the KK-expansion -- both from the 10d metric and other bulk fields. One can convince oneself that if all those 4d fields are canonically normalized, then the suppression by $1/M_4$ is a universal feature of all 3-vertices. This is a key observation: It implies that all 3-vertex-based loop corrections to the massless 4d graviton or volume modulus propagator (see l.h.~side of Fig.~\ref{loopdiagrams}) have the same parametric behavior. This holds independently of the kind of field running in the loop. Other 1-loop contributions come from tadpole diagrams involving a 4-vertex (see~r.h.~side of Fig.~\ref{loopdiagrams}). The 4-vertices are universally suppressed by $1/M_4^2$, such that all tadpole diagrams have the same parametric behavior as the loops built with two 3-vertices.

Using the KK action of \eqref{kkreduction} and the diagrams in Fig.~\ref{loopdiagrams}, one can in principle explicitly compute the 1-loop correction to the propagators of the massless 4d graviton $h_{\mu \nu}^{0}$ and the volume modulus. Let us first focus on the graviton correction. It can be interpreted as a correction $\delta_{R_4}$ to the Ricci scalar term in the Einstein-frame action:
\be
\left(S+\Delta S\right)_\text{E} = \frac{M_{4}^2}{2}\int \dd^4x \sqrt{-g} \left( 1 + \delta_{R_4}\right) R_4+\cdots \,.
\ee
In dimensional regularization, the loop contribution takes the form
\begin{equation}
\delta_{R_4}^\varepsilon = \left. \frac{\dd}{\dd p^2} \right|_{p^2=0} \kappa_{4}^2 \mu^\varepsilon \sum\limits_{a} \int\dd^{4-\varepsilon}q \frac{f_4\left(p,q,m_{a}\right)}{\left( q^2+ m_{a}^2\right) \left((p-q)^2 + m_{a}^2\right)},
\label{dimreg}
\end{equation}
where $f_4\left( p,q,m_{a} \right)$ is of mass dimension $4$. For the 3-vertex contribution, this follows from the fact that each 3-vertex comes with two derivatives. For the 4-vertex contribution, one has two derivatives from the vertex and a term cancelling the $(p-q)^2$-expression in the denominator. The sum over KK modes gives \eqref{dimreg} a maximal degree of divergence which is as strong as in 10d, i.e.~octic in cut-off language. The final 4d correction is obtained as
\be
\delta_{R_4}=\lim_{\varepsilon\to 0} \left(\delta_{R_4}^\varepsilon+\delta_{R_4}^{\varepsilon,\,{\rm c.t.}}\right)\,,
\label{deltaR}
\ee
i.e.~after adding the counterterm contribution and taking $\varepsilon$ to zero. To be specific, we use minimal subtraction, such that $\delta_{R_4}^{\varepsilon,\,{\rm c.t.}}=\,\mbox{const.}/\varepsilon$.

If the integral in \eqref{dimreg} were finite, and hence no counterterm were needed, then on dimensional grounds one would find
\be
\delta_{R_4}=\frac{\mathcal{O}(1)}{L^2M_{4}^2}\,.
\label{finitepiece}
\ee
This follows because $m_{a}^2=f_{a}/L^2$, with $f_{a}$ dimensionless numbers encoding the Calabi-Yau geometry.
Then, the loop correction in Einstein frame takes the form
	\begin{equation}
		\left(S+\Delta S\right)_\text{E} = \frac{M_{4}^2}{2}\int \dd^4x \sqrt{-g} \left( 1 + \delta_{R_4}\right) R_4=\frac{M_{4}^2}{2}\int \dd^4x \sqrt{-g} \left( 1 + \frac{\mathcal{O}(1)}{L^2M_{4}^2}\right) R_4\,,
		\label{gravpropcorr}
	\end{equation}
in agreement with the first term in \eqref{sloopcorr}.

Let us now discuss the precise form of \eqref{deltaR}. Note that a non-zero counterterm $\delta_{ R_4}^{\varepsilon,\,{\rm c.t.}}$ comes with a pole in \eqref{dimreg} and the latter is necessarily accompanied by a factor $(\mu L)^\varepsilon$. In the limit $\varepsilon\to 0$, a finite term $\sim \ln(\mu L)$ is left. If we take our theory to be defined at the string scale, we may set $\mu=M_s=M_{10}g_s^{1/4}$ and the resulting logarithm would represent a significant enhancement of the ${\cal O}(1)$ coefficient in \eqref{finitepiece}.

In the following we will argue that such terms do not occur. The reason is that $\delta_{ R_4}^{\varepsilon,\,{\rm c.t.}}$ vanishes. To understand this, note that compactifying of a theory on a smooth manifold represents an IR modification and does not affect the UV structure. Hence all 4d counterterms derive from 10d counterterms. It is clear that a counterterm proportional to $R_{10}$ in 10d will induce a counterterm proportional to $R_4$ after compactification. However, this is not the only option. For instance, if the 10d action contains a term of the form\footnote{We denote by $R_{10}^n$ any $n$th power term in the Riemann tensor with all indices contracted.} $R_{10}^5$, one way of compactifying this term is schematically as $R_{\text{external}}\int_{\mathcal{M}_6}d^6 y R^4_{\text{internal}}$. A 10d counterterm $\sim R^5_{10}$, which is completely unrelated to the 10d propagator of $h_{MN}$, can hence induce a 4d counterterm relevant for the propagator of $h_{\mu \nu}$ and thereby signal a logarithmic enhancement of $\delta_{R_4}$.

To study 4d counterterms we then require some information about higher-order terms in the 10d action. It is known that there are no terms of order $R_{10}^2$, $R_{10}^3$ \cite{Green:1987mn}, or $R_{10}^5$  \cite{Richards:2008jg} in the IIB supergravity action.

First let us consider the 1-loop correction to $R_{10}$. By analogy to \eqref{dimreg}, it takes the form
	\begin{equation}
		\delta_{{R}_{10}}^\varepsilon= \left. \frac{\dd}{\dd p^2} \right|_{p^2=0} \kappa_{10}^2 \mu^\varepsilon \int\dd^{10-\varepsilon}q \frac{f_{10}(p,q)}{ q^2(p-q)^2 },
		\label{dimreg10d}
	\end{equation}
where $f_{10}$ is again a function of mass dimension $4$. The dominant divergence, taking into account the $p^2$ derivative, is octic. Nevertheless, one could in principle imagine a sub-leading logarithmic divergence and hence a pole being present.\footnote{
As 
a simple example where subleading logarithmic corrections occur, consider the 1-loop correction to the Higgs mass:
\begin{equation}
	\sim \frac{\lambda}{2} \int \frac{\dd^4p}{(2\pi)^4} \frac{1}{p^2+m^2} = -\frac{\lambda m^2}{32\pi^2}\left( \frac{\Lambda^2}{m^2}-\ln\frac{\Lambda^2}{m^2} \right)+\mathcal{O}(\Lambda^0)\,.
\end{equation}
In $4-\varepsilon$ dimensions, one would find a pole proportional to $m^2$.
}
However, here this can be excluded on dimensional grounds since no dimensionful parameters like a mass appear and $p^2$ is set to zero. Thus, $\delta_{{R}_{10}}^\varepsilon$ vanishes.

This argument can be repeated word by word for the $R_{10}^4$ term: The dominant 1-loop divergence for its coefficient is quadratic and, since no mass scale is available, there is no sub-leading logarithmic divergence. Hence there is no pole and no non-zero counterterm arises.

Let us make a side remark concerning specifically the $R_{10}^4$ term (but also relevant more generally): This higher-order term contributes a 4-vertex and can hence also correct the propagator via a diagram of the form Fig.~\ref{loopdiagrams} (b). The loop diagram with this $R_{10}^4$ vertex is, however, suppressed by a factor $M_{10}^{6}$ compared to the analogous loop with a 4-vertex from $R_{10}$. We may hence neglect it.

Note that if the ten-dimensional action would have included a term of order $R^5_{10}$, such a term could have produced a pole in 10d since it is a marginal operator. This could in turn have produced a logarithmic term for the four-dimensional propagator\footnote{This happens for example in four-dimensional higher derivative gravity with a marginal $R^2$ term \cite{Donoghue:1994dn}.}.
The absence of such a term in ten dimensions combined with the absence of counterterms in 1-loop diagrams constructed from $R_{10}$ and $R_{10}^4$ 
ensures that there are no counterterms in ten dimensions that could induce a logarithmic term in four dimensions.

As \eqref{sloopcorr} suggests, there are also direct corrections to the volume modulus kinetic term coming from the loop diagrams in Fig.~\ref{loopdiagrams}, with the external legs belonging to the modulus. If the modulus is canonically normalized, the 3- and 4-vertices are again suppressed by $1/M_{4}$ and $1/M_4^2$. This leads to the same form of loop integrals as for $R_4$. After returning to the non-canonical field $L$, one finds
	\begin{equation}
		M_{4}^2\int\dd^4x\, \frac{\mathcal{O}(1)}{L^2M_{4}^2}\,\, \frac{\left( \partial L \right)^2}{L^2}\,,
	\end{equation}
consistently with the correction proportional to $b_1$ in \eqref{sloopcorr}.

In summary, the vertices all scale the same way, regardless of the fields correcting the propagator: 3-vertices are suppressed by $1/M_{4}$ and 4-vertices by $1/M_{4}^2$. This results in the universal form of the 1-loop correction proposed in \eqref{efaction} and \cite{vonGersdorff:2005bf}. The leading correction to the \K potential stemming from genuine loop corrections is therefore proportional to $1/\tau^2$, with $\tau$ the 4-cycle volume. The absence of a logarithmic enhancement is a non-trivial consequence of the divergence structure of the 10d effective supergravity theory.

%%%%%%%%%%%%%%%%%%%%%%%%%%%%%
\subsection{Local $\alpha'$ Corrections from the Bulk Theory} \label{localalphacorrections}
%%%%%%%%%%%%%%%%%%%%%%%%%%%%%

Our field-theoretic loop analysis of the last subsection required the discussion of 10d higher-curvature terms, which are needed to absorb UV divergences. Specifically, the absence of an $R_{10}^5$ term prevented the appearance of a logarithmic correction in 4d. In this context, it may be useful to provide a short, more general discussion of such higher-curvature terms and the resulting corrections to the \K potential (the local $\alpha'$ corrections). Our line of reasoning will be directly applicable to similar higher-curvature terms localized on branes, O-planes and their intersections, where the implications are less well-established and hence more interesting. From now on we will explicitly keep track of $g_s$. In a purely low-energy EFT perspective, it can be understood as $g_s\sim \Lambda^4/M_{10}^4$, where $\Lambda=M_s$ is the EFT cutoff.

We focus on the purely gravitational (higher) curvature part of the type-IIB action in the Einstein frame. Suppressing numerical prefactors for brevity, it reads \cite{Antoniadis:1997eg}
\begin{equation}
    S_\text{EF} \sim \int \dd^{10}x \sqrt{-g} \bigg[ M_{10}^{8} R_{10} + \frac{M_{10}^{2}}{g_s^{3/2}}R_{10}^4  + M_{10}^2g_s^{1/2} R_{10}^4 +\mathcal{O}\left(M_{10}^{-2}g_s^{-5/2}R_{10}^6 \right) \bigg].
    \label{schematic10daction1}
\end{equation}
The first two terms appear at string tree-level: the Einstein-Hilbert term and the $R_{10}^4$ correction. The third term arises at string-theoretic 1-loop order \cite{Antoniadis:1997eg} and hence comes with a relative $g_s^2$ suppression. In our field-theoretic approach, its size is set by the quadratic divergence of the $R_{10}^4$ term, such that its coefficient can also be understood as $\Lambda^2$. In other words, part of this term may be identified as a counterterm of the EFT analysis. Note that the Einstein-Hilbert term does not receive a string-theoretic 1-loop correction \cite{Polchinski:1998rr}. We will not discuss corrections at order $R_{10}^6$ since their effects on the \K potential are subleading compared to genuine loop effects.

Independently of the genuine loop effects, higher-curvature terms affect the 4d Einstein-Hilbert term obtained after compactification. Specifically, 
dimensionally reducing the $R^4_{10}$ term as
\begin{equation}
\left(\frac{M^{2}_{10}}{g_s^{3/2}}+M_{10}^2g_s^{1/2}\right)R_{\text{external}} \int \dd^6 x R_{\text{internal}}^{3}\sim \left(\frac{M^{2}_{10}}{g_s^{3/2}}+M_{10}^2g_s^{1/2}\right) R_{\text{external}}
\end{equation}
reproduces the well known string tree-level BBHL correction \cite{Antoniadis:1997eg, Becker:2002nn} and its 1-loop counterpart \cite{Antoniadis:1997eg}. Comparing to the tree-level term $M_{10}^8 L^6 R_{\rm external}$, we see that their relative size is $1/(M_{10}^6 L^6 g_s^{3/2})$ and $g_s^{1/2}/(M_{10}^6 L^6)$ respectively. This is also the scaling of the corresponding corrections to the \K potential, which arise after Weyl rescaling to 4d Einstein frame.

%%%%%%%%%%%%%%%%%%%%%%%%%%%%%
%%%%%%%%%%%%%%%%%%%%%%%%%%%%%
\section{Extending and Generalizing the Basic Analysis} \label{generalization}
%%%%%%%%%%%%%%%%%%%%%%%%%%%%%
%%%%%%%%%%%%%%%%%%%%%%%%%%%%%

%%%%%%%%%%%%%%%%%%%%%%%%%%%%%
\subsection{D-brane and O-plane corrections}
\label{membranesection}
%%%%%%%%%%%%%%%%%%%%%%%%%%%%%

We still need to consider additional corrections due to extended objects filling the four external spacetime dimensions. We will focus on D3/D7-branes and O3/O7-planes as they are relevant to phenomenological compactifications such as the LVS, but the same logic applies to other extended objects, like for example even dimensional D-branes/O-planes of type IIA. To study corrections induced by D-branes, we follow the procedure of Sect.~\ref{feynman}. However, the KK tower of 10d type-IIB bulk fields is now replaced by the analogous tower resulting from the compactification of the worldvolume theory on the brane. For O-planes no such additional tower exists, but the bulk-field KK tower and hence the corresponding loop correction is modified by the orientifold projection. Both for D-branes and O-planes new operators localized on the brane, on intersection cycles, or at the singularity potentially come into play.

Our D$p$-branes/O$p$-planes wrap $p-3$ cycles in the internal dimensions. For the moment, we assume our compact geometry to be governed by a single length scale $L$. This is then also the typical length scale of these $p-3$ cycles. Further down in this section we will also comment on generalizations to cases with multiple Kahler moduli. Scenarios with hierarchically different cycles will be considered in Sects.~\ref{dimblowup} and~\ref{fibreinflation}.

%%%%%%%%%%%%%%%%%%%%%%%%%%%%%
\subsubsection{D-branes} \label{dbranes}
%%%%%%%%%%%%%%%%%%%%%%%%%%%%%

The fields of the gauge multiplet living on the brane couple to the graviton and its moduli. Hence, these fields run in loops, such as in Fig. \ref{loopdiagrams}. It is easy to see that, as long as $L$ remains the only relevant length scale, the couplings to graviton and moduli still come with factors $1/M_4$. More precisely, 3-vertices and 4-vertices are again universally suppressed by $1/M_4$ and $1/M_4^2$, respectively. Therefore, the new genuine loop corrections are parametrically the same as those computed in Sect.~\ref{feynman}.

In addition, we should consider $R_{p+1}^n$ terms on the brane. In analogy to our previous discussion of $R_{10}^n$ terms, brane localised higher-curvature terms can, after dimensional reduction, impact the coefficient of $R_4$. 
Again, two distinct effects arise:

First, if the $R_{p+1}^n$ term is marginal, its coefficient at one-loop order can contain a $1/\varepsilon$ counterterm. This means that the one-loop correction from the brane KK-tower can contain a corresponding $1/\varepsilon$ pole. Hence, a logarithm $\ln(M_{10}g_s^{1/4}L)$ can appear in the loop correction to the coefficient of $R_4$. In terms of our classification proposed in the Introduction, this effect is on the boundary between a localized $\alpha'$ correction and a genuine loop correction. For definiteness, we will count it as part of localized $\alpha'$ corrections. This appears sensible since a logarithmic integral over momentum scales $\mu$ in the range between $1/L$ and $\Lambda=M_{10}g_s^{1/4}$ is dominated by scales which satisfy $\mu\gg 1/L$. The effect is hence localized in the sense of not being sensitive to the non-trivial CY geometry with typical length scale $L$. 

Second, if the $R_{p+1}^n$ operator is relevant, the coefficient includes possible power-like divergences, cut off at the string scale (see Sect.~\ref{localalphacorrections}). From the perspective of a loop calculation, such operators generically supply counterterms, which however happen to vanish in dimensional regularization. Thus, for us only the classical part of the coefficient is relevant, providing a localized $\alpha'$ correction.

We will only consider $R_{p+1}^n$ terms up to and including $n=(p+1)/2$, which corresponds to the operator being marginal. Irrelevant operators will not contribute at the same order in $1/L$ as the loop effects we are interested in.

The worldvolume action of a $p$-brane contains two types of curvature corrections. Firstly, there are curvature corrections to the DBI action \cite{Bachas:1999um}. For further work, including curvature-gauge-field terms, see e.g.~\cite{Wijnholt:2003pw,Garousi:2009dj,Becker:2010ij,Jafari:2016vbr,Garousi:2017fbe,Akou:2020mxx}. The curvature corrections start at order $R_{p+1}^2$. Higher order terms in $R_{p+1}$ have to our knowledge not been computed, but we expect that such terms exist. We will hence include them in our discussion, with the caveat that some of them may turn out to be forbidden. Suppressing again numerical prefactors, the curvature corrections to the DBI action then take the form\footnote{In our understanding, it is expected \cite{Bachas:1999um,Wyllard:2000qe,Fotopoulos:2001pt} that there is no $R^3$ term but the $R^4$ term is present.}
\begin{equation}
    S_{\text{DBI}} \supset \int \dd^{p+1}x \left[ M_{10}^{p-3}g_s^{(p-7)/4}(1+g_s) R_{p+1}^2 + M_{10}^{p-5}g_s^{(p-9)/4} (1+g_s)R_{p+1}^3 + .. \right]\,.
    \label{DBIactionschematic}
\end{equation}
This equation is written in Einstein frame in the sense that the varying part of the dilaton is absorbed into the metric. The $g_s$ scaling follows from the substitution $M_s = \Lambda= M_{10} g_s^{1/4}$. For each of the operators in \eqref{DBIactionschematic}, the second, $g_s$-suppressed term in the round bracket can be interpreted as coming from a power-like divergence cut off at the string scale.

In addition, there are topological curvature terms in the Wess-Zumino (WZ) action. Their form is known, see e.g.~\cite{Green:1996dd, Bachas:1999um}. These terms consist of couplings between the bulk Ramond-Ramond fields and even powers of the curvature two-form. Since our analysis is focused on backgrounds without fluxes, we will not consider these terms here. However, even in the absence of background fluxes, D3/O3 loci source $F_5$ while D7/O7 loci source $F_9$ (and possibly an induced $F_5$ field strength). For D7/O7, the induced $F_9$ RR-field strength can be set to zero by cancelling the D7 tadpole locally. For D3/O3, local D3 tadpole cancellation can not be achieved. 

The resulting $F_5$ field strength is a source for warping to be discussed in Sect.~\ref{warping}. In fact, it is well known since \cite{Giddings:2001yu} that the leading-order $F_5$ background is fixed together with the warp factor.  
As soon as higher-order corrections like, for example, higher-order terms in the WZ action are included, $F_5$ effects may become relevant independently of warping, but this would correspond to a superposition of classical backreaction and higher-order $\alpha'$ effects. Hence, this goes beyond the goals of the present paper, where we limit our discussion to each effect separately.

Consider first the impact of counterterms from marginal operators in \eqref{DBIactionschematic} that signal the possibility of a logarithmic term in the $R_4$ coefficient.

For the D3-branes, the leading $R_4^2$ term is marginal. There is then potentially a $1/\varepsilon$ counterterm and hence a logarithmic enhancement. However, this cannot possibly impact the $R_4$ term as the D3-brane is pointlike in the internal dimensions and there is no dimensional reduction to be done that could turn the $R_4^2$ term into an $R_4$ term. Moreover, similarly to our discussion in the case of $R_{10}^4$ in the bulk, a contribution of $R_4^2$ via the induced vertex will be subleading.

Concerning the D7-branes, a possible $R_8^4$ term would be a marginal operator. This may give rise to a $1/\varepsilon$ counterterm, leading to a logarithmic enhancement of the form $\ln(M_{10}g_s^{1/4}L)$ in the coefficient of $R_4$. We note that the logic here is exactly the same as for a potential $R^{5}_{10}$ bulk term in Sect.~\ref{feynman}, which is however known to be absent. The expected appearance of a logarithmic term, related to D7-branes, in our analysis of loop corrections is extremely interesting: If present, it would be dominant compared to genuine loop effects. Moreover, it is conceivable that the numerical coefficient of this logarithm is calculable since it is a universal feature of the UV structure of the 10d theory with D7-branes in flat space.

In the context of logarithmic corrections, let us comment on the log effect at order $\alpha'^3$ analysed in \cite{Antoniadis:2018hqy}, which may also be used for the construction of novel (A)dS vacua. The relevant correction arises in torus orbifolds from the combination of two effects: First, there is the $R^4$ term in the bulk (the third term in \eqref{schematic10daction1}), which is localized at the points of high curvature.\footnote{We note in passing that this localization as well as the absence of a corresponding tree-level term is a special feature of torus orbifold models.} Second, there is a backreaction on the $R^4$ term sourced, in this case, by a D7-brane. The logarithm comes from the codimension-2 behaviour of the relevant Greens function. Crucially, this backreaction is claimed not to involve a further $\alpha'$ suppression, possibly related to the assumption that D7-tadpoles are not cancelled locally, i.e.~O7/D7 branes do not come in $SO(8)$ stacks. We note that the log term of \cite{Antoniadis:2018hqy} appears at the order $\alpha'^3$, while our previously discussed logarithm arises at the order of genuine 1-loop effects, i.e.~$\alpha'^4$. The correction of \cite{Antoniadis:2018hqy} can be understood as a combination of a local $\alpha'$ correction, the $R^4$ part, and a warping effect, the backreaction of the D7 brane on the geometry where the curvature is localized. We will discuss warping in detail in Sects.~\ref{warping} and \ref{comparison}. A similar correction, based on the interplay of warping and higher-curvature terms, has been recently discussed in \cite{Junghans:2022exo, Gao:2022fdi}. There, the warping does not come from D7-branes and the effect arises at higher order in $\alpha'$. In the present paper, we do not consider corrections which need warping and higher-curvature terms at the same time. Clearly, such formally higher-order effects can nevertheless be important and should be systematically studied in the future.

Consider now the local $\alpha'$ corrections of branes to the $R_4$ coupling.

D3-branes are pointlike in the internal dimensions, so no dimensional reduction has to be done.  There is then no way for the leading $R_4^2$ term in \eqref{DBIactionschematic} or any higher term $R_4^n$ to contribute to the coefficient of $R_4$ in the 4d EFT.

D7-branes wrap 4-cycles with typical length scale $L$. Thus, $R_8^n$ terms contribute as 
\begin{equation}
\begin{split}
M^{8-2n}_{10}g_s^{(2-n)/2}\left(1+g_s\right)R_{\text{external}}& \int_{4-\text{cycle}} \dd^4 y\, R_{\text{internal}}^{n-1}\\ & \sim M^{8-2n}_{10}g_s^{(2-n)/2}\left(1+g_s\right) L^{4-2(n-1)} R_{\text{external}}\,,
\end{split}
\label{classicalcontribution}
\end{equation}
with the leading term arising for $n=2$. The effects of such an $R_8^2$ term have been studied in \cite{Grimm:2013gma,Grimm:2013bha,Junghans:2014zla,Weissenbacher:2019mef,Weissenbacher:2020cyf}. The $M_{10}^4R^2$ term induces a field redefinition and does not correct the \K potential \cite{Grimm:2013bha,Grimm:2013gma,Junghans:2014zla}. However, as displayed in \eqref{classicalcontribution}, a subleading term $M_{10}^4g_sR^2$ may in general be present and induce a correction to the \K potential.
Its presence has to our knowledge not yet been confirmed by string amplitude calculations. Such a term would contribute at order $M_{10}^4 L^2 g_s$ to $R_4$. This would lead to a correction proportional to $g_s$ and of degree $-1$ in 4-cycles to the \K potential, which is dominant compared to BBHL \cite{Becker:2002nn}. We note that this matches the KK correction of the BHP conjecture. At the level of the scalar potential, this correction will be subleading compared to BBHL but of the order of genuine loop corrections due to the extended no-scale structure \cite{vonGersdorff:2005bf, Berg:2007wt, Cicoli:2007xp}. 

If an $R_8^3$ term in \eqref{DBIactionschematic} should exist, it would via \eqref{classicalcontribution} contribute at the order of BBHL \cite{Antoniadis:1997eg, Becker:2002nn}. Since it is not subject to an extended no-scale cancellation, this is dominant compared to loop effects on the level of the scalar potential. Even though we have so far not discussed the case of multiple \K moduli, let us briefly note that an $R_8^3$ would be particularly interesting in this context.
Dimensionally reducing the term as above gives
\begin{equation}
    \frac{M_{10}^2}{g_s^{1/2}}\int\dd^8x R_8^3 \sim \frac{M_{10}^2}{g_s^{1/2}}\int \dd^4 y R_\text{internal}^2 \int\dd^4x R_4 \sim \frac{M_{10}^2}{g_s^{1/2}}f(\tau_1,\dots,\tau_n) \int\dd^4x R_4\,,
\end{equation}
where $n$ labels the \K moduli $\tau_i$ and  $f(\tau_1,\dots,\tau_n)$ is a homogeneous function of degree $0$. Crucially, it is possible that $f$ is not just a constant but depends non-trivially on the ratios of 4-cycles. It could hence be the dominant effect lifting the flat directions associated with `large' 4-cycle ratios, as it is typically required in the LVS. 

Finally, the dimensional reduction of an $R_8^4$ term in \eqref{classicalcontribution} would result in a correction comparable to one-loop effects, but without logarithmic enhancement.

%%%%%%%%%%%%%%%%%%%%%%%%%%%%%
\subsubsection{O-planes}
%%%%%%%%%%%%%%%%%%%%%%%%%%%%%

In contrast to D-branes, O-planes do not come with new fields propagating on their world-volume. Thus, no new contributions to the diagrams of Fig.~\ref{loopdiagrams} arise.

The curvature terms of the type of  \eqref{DBIactionschematic} also exists for O-planes. At the order $\alpha'^2$, where the corrections are known, the $R^2$ curvature term on the O-plane is $2^{p-5}$ times that on the D-brane. Crucially, D-brane and O-plane curvature correction have the same sign, so they do not cancel against each other. The dimensional reduction of curvature terms on the O-plane worldvolume action then proceeds entirely analogously to the D-brane case.

O-planes have two further effects that are not present for D-branes.
First, the orientifold projection removes part of the KK modes. Thus, the KK spectrum relevant for the 4d action \eqref{kkreduction} is modified. The parametric form of the resulting loop correction remains, however, unchanged. Note also that local 10d physics away from the orientifold plane is not affected by the projection, such that the analysis of 10d divergences and counterterms goes through as before.

Second, the orientifolding changes the geometry of the compact space in a UV sensitive manner. Put differently, the O-plane hypersurface represents a singularity within the surrounding, weakly curved 10d geometry. Thus, our logic in Sect.~\ref{feynman}, which assumed that the UV structure and in particular the counterterms are those of the flat 10d theory, does not apply any more. Instead, the loop calculation involving the orientifold-projected KK spectrum may require counterterms localized at the O-plane. These are  the same operators that we discussed above as possible curvature corrections on O-planes and D-branes. Thus, no entirely new effects arise and our previous discussion of loop corrections from the bulk and from D-branes, including the possible log-enhancements, remains valid. Crucially, after an orientifold projection introducing O7-planes a second source for log-enhancements which we discussed in the D7-brane context appears: It is due to the projected spectrum of bulk modes, which may induce a log-divergent $R_8^4$ term on the O7-plane.

\subsubsection{Intersecting D-branes and O-planes} \label{intersectingbranes}

Finally, let us discuss setups where D-branes/O-planes intersect. We focus again on type-IIB orientifolds with D3/D7-branes and O3/O7-planes. In this setting, only D7/O7 intersections are relevant, filling out curves in the internal space. We assume that their 2d geometry is governed by a single length scale $L$. In total, the intersection manifold is 6-dimensional and potentially supports new operators. We focus on curvature effects, neglecting fluxes and couplings of the branes to higher form fields.

Fields living on the intersection couple to the graviton and its moduli, hence inducing loop correction as in Fig.~\ref{loopdiagrams}. The 3- and 4-vertices are again universally suppressed by $1/M_4$ and $1/M_4^2$, respectively. This leads to the same parametric behavior of genuine loop corrections as observed before.

As usual, UV divergences of loop corrections are absorbed in local operators, the most interesting being the marginal operator $R^3$. If this operator is allowed and the corresponding divergence arises, a logarithmic enhancement in the coefficient of $R_4$ is induced. 

Of the other local curvature operators, the most import one is the Einstein-Hilbert term:
\begin{equation}
    S_{\text{int,EH}}\sim M_{10}^4 (1+g_s)\int\dd^6x \,R_6\,.
    \label{sintersection}
\end{equation}
Here we have displayed both the tree level and the string-one-loop contribution. 
The tree level term does not lead to a correction of the \K potential but only to a field redefinition \cite{Grimm:2013bha,Grimm:2013gma,Junghans:2014zla}.
This is supported by scattering analyses in type IIA on intersecting D6-branes/O6-planes\footnote{It would very interesting to study the effect of this term in the context of DGKT \cite{DeWolfe:2005uu}.} \cite{Epple:2004ra} and in type IIB with D9/D5-branes \cite{Haack:2015pbv}. They show that an Einstein-Hilbert term on brane intersections can only be induced at 1-loop level, corresponding to the $g_s$-suppressed term in \eqref{sintersection}.
References \cite{Epple:2004ra,Haack:2015pbv} discuss the contribution of this term to $R_4$, which is of order $M_{10}^4 L^2g_s$:
\begin{equation}
        S_{\text{int,EH}}\sim M_{10}^4g_s\int\dd^6x R_6 \sim M_{10}^4g_s L^2 \int\dd^4x R_4\,. 
        \label{Sintersection}
\end{equation}
This matches the EFT analysis of \cite{Junghans:2014zla}. In this analysis, one starts from the string-frame $R^2_8$ operator on D7/O7. Taking into account the Weyl rescaling to the 10d Einstein frame together with the varying dilaton near D7-branes, one of the Ricci scalars may be replaced by dilaton gradients. One is then left with an integral over the remaining Ricci scalar which is effectively localized on the D7/O7 intersection. This localization is due to the non-trivial dilaton profile which one brane induces in the vicinity of the intersecting brane. The net effect is an $R_6$ operator on the brane intersection, to be viewed as a local $\alpha'$ correction.
The resulting correction to the \K potential is proportional to $g_s$ and of degree $-1$ in 4-cycles and its effect on the scalar potential is subject to the extended no-scale structure. 
Not much is known about higher order operators on the intersection cycle such as $R_6^2$ and $R_6^3$. A comment on this issue can be found in \cite{Cicoli:2021rub}.
If terms of the form $R_6^2$ and $R_6^3$ on the intersection locus exist, they would induce correction with the volume-scaling of BBHL (but suppressed in $g_s$) and of genuine loops effects respectively.

Let us briefly comment on the possible $R_6^2$ term in more detail. Using the metric ansatz \eqref{ansatz}, the $R_6^2$ term contributes to the 4d Einstein-Hilbert term through the following dimensional reduction:
\begin{equation}
    \frac{M_{10}^2}{g_s^{1/2}}\int\dd^6x R_6^2 \sim  \frac{M_{10}^2}{g_s^{1/2}} \int \dd^2 y \,R_{\rm internal} 
    \int\dd^4x R_4
    \sim 
    \frac{M_{10}^2}{g_s^{1/2}} \chi(S)
    \int\dd^4x R_4\,.
\end{equation}
Here $\chi(S)$ is the Euler characteristic of the intersection surface $S$.
Comparing this with the tree-level term $M_{10}^8 L^6 R_4$, we see that the relative, parametric suppression of the correction from an $R_6^2$ term on a 7-brane intersection locus is $1/(L^6M_{10}^6g_s^{1/2})\sim g_s(l_s^6/L^6)$. This is down by a factor $g_s$ compared to BBHL.

%%%%%%%%%%%%%%%%%%%%%%%%%%%%%
\subsubsection{Summary}
%%%%%%%%%%%%%%%%%%%%%%%%%%%%%

In this section we have studied brane-induced corrections to the four-dimensional Kahler potential. Our goal was to demonstrate that branes do not spoil the analysis of Sects.~\ref{dimsingle} and \ref{feynman}. We have seen that, indeed, brane effects do not alter the power of the volume with which genuine loop corrections scale. However, in the presence of D7-branes/O7-planes, log-enhanced terms may arise. They are expected to be dominant since $M_{10}g_s^{1/4}L\gg 1$ and hence, though to lesser extent, also $\ln(M_{10}g_s^{1/4}L)\gg 1$. The log-enhanced contribution then wins against the $\mathcal{O}(1)$ numerical coefficient $b_0$ in \eqref{sloopcorr}. This correction would then be decisive for all moduli stabilization schemes relying on loop corrections. The marginal operators potentially responsible for this effect are of type $R_8^4$ for 7-branes and of type $R_6^3$ for their intersections. It would therefore be very important to know whether these terms are really present and to determine their coefficients.

In setups with intersecting D7-branes and/or O7-planes, it has been shown that an Einstein-Hilbert term localized on the intersection curve is induced at 1-loop level. Local $\alpha'$ corrections coming from this operator then lead to corrections to the \K potential proportional to $g_s$ and of degree $-1$ in 4-cycles. This is fundamentally different from the genuine loop corrections of degree $-2$ in 4-cycles. Terms of degree $-1$ in 4-cycles can also be obtained from an $M_{10}^2g_sR_8^2$ operator on a D7/O7. 

In cases with multiple Kahler moduli the corrections considered in this section can be even more interesting since ratios of 4-cycles can potentially appear. These ratios can be large given a hierarchical structure in the Kahler moduli. An explicit example where large ratios appear is discussed in Sect.~\ref{inverse}.

We emphasize once again that for some of the corrections discussed it is not yet clear whether the required term really appears in the DBI action and whether its dimensional reduction works as displayed schematically in  \eqref{classicalcontribution}. Moreover, one needs to understand whether the resulting effect can be absorbed in a field redefiniton.\footnote{By this we mean that the Kahler manifold as an abstract mathematical object remains unchanged, only the coordinates are modified. In other words, a given point on this manifold might change its interpretation in terms of the volumes of some set of 4-cycles, measured in string units. This implies that, from the perspective of the 4d supergravity model, there is no change (as long as the above 4-cycles do not enter the model in some other way, e.g. through the non-perturbative superpotential). An observer having access to the `microscopic information' of 4-cycle volumes in string units could discover the correction.}

%%%%%%%%%%%%%%%%%%%%%%%%%%%%%
\subsection{Multiple Kahler Moduli} \label{dimmulti}
%%%%%%%%%%%%%%%%%%%%%%%%%%%%%

Most Calabi-Yau manifolds have more than a single \K modulus. Moreover, the LVS requires at least two \K moduli. It is therefore crucial to extend the analysis above to Calabi-Yaus with multiple \K moduli. This is the goal of the present subsection. The fundamental result is the same as in the single-modulus case: The genuine loop correction to the Kahler potential is a homogeneous function of degree $-2$ in 4-cycle volumes. A logarithmic enhancement is again possible. Readers who are prepared to accept these facts may skip to Sect.~\ref{warping}.

To demonstrate our claims, let us first recall some basics concerning the \K moduli sector of type-IIB orientifolds with D3/D7-branes. The tree-level \K potential $K$ and volume $\mathcal{V}$ of the internal manifold $\mathcal{M}_6$ read
\begin{equation}
    K=-2\ln(\mathcal{V}),~~~~~~~~\mathcal{V} = \frac{1}{3!} \int_{\mathcal{M}_6} J \wedge J \wedge J = \frac{1}{3!} \mathcal{K}_{ijk} t^i t^j t^k\,,
\end{equation}
with $J$ the Kahler form and $\mathcal{K}_{ijk}$ the triple intersection numbers. The two-cycle \K moduli $t^i$ are related to four-cycle \K moduli $\tau_i$ as 
\begin{equation}
    \tau_i = \frac{\partial \mathcal{V}}{\partial t^i} = \frac{1}{2} \mathcal{K}_{ijk} t^j t^k\,.
\end{equation}
The $\tau_i$ ($t^i$) measure the Einstein frame 4-cycle (2-cycle) volume in units of $l_s=2\pi\sqrt{\alpha'}$. The \K potential $K$ has to be interpreted as a function of the complexified 4-cycle moduli $T_i$. This is achieved by expressing the $t^i$ through the $\tau_i$ and the latter as $\tau_i=(T_i+\overline{T}_{\overline{\imath}})/2$.

To argue for the parametric form of loop corrections, we introduce dimensionful \K moduli as follows:
    \begin{equation}
        \tilde{t}^i= \frac{t^i}{M_{10}^2}\,,~~~~~~~~\tilde{\tau}_i = \frac{\tau_i}{M_{10}^4}\,,~~~~~~~~ \tilde{\mathcal{V}}=\frac{\mathcal{V}}{M_{10}^6}\,.
    \end{equation}
The dimensionful quantities are characterized by a tilde.

Let us start from the 10D IIB action \eqref{2baction} and the metric ansatz \eqref{ansatz} (but now with multiple Kahler moduli) and dimensionally reduce to four dimensions. This yields the four-dimensional Jordan frame action. At tree level we have\footnote{From here on we change our index conventions slightly: We use the 4-cycle-moduli as coordinates on the moduli space, hence giving them upper indices.}
\begin{equation}
    S_{\text{JBD}} = \frac{1}{2\kappa_{10}^2}\int\dd^4x \sqrt{-g}\tilde{\mathcal{V}}\left[ R_4+ \tilde{F}_{ij}\partial_\mu\tilde{\tau}^i\partial^\mu\tilde{\tau}^j+\cdots \right]\,.
    \label{multiSJBD}
\end{equation}
Here we display only the Einstein-Hilbert term and kinetic terms of the dimensionful 4-cycle moduli $\tilde{\tau}^i$, with $\tilde{F}_{ij}$ denoting their prefactors.\footnote{
After Weyl rescaling to the Einstein frame, these prefactors take the form of second derivatives of the tree-level \K potential (see e.g.~\cite{Bodner:1990zm} for the corresponding 2-cycle calculation and \cite{Becker:2002nn,Grimm:2004uq} for the transition to 4-cycles).
}
One-loop corrections come from integrating out the tower of KK modes and from the fluctuations of the moduli themselves. Exactly as in the single-modulus analysis of Sect.~\ref{dimsingle}, the UV-scale $M_{10}$ can not appear in the result, except through divergences associated with higher-dimension operators in 10d or on branes (cf.~Sects.~\ref{feynman} and~\ref{membranesection}). Thus, on dimensional grounds one expects
    \begin{equation}
        \Delta S_{\text{JBD}} = \int \dd^4x \sqrt{-g} \left( \tilde{a}(\{\tilde{\tau}_k\})R_4+\tilde{b}_{ij}(\{\tilde{\tau}_k\})\partial_\mu\tilde{\tau}^i\partial^\mu\tilde{\tau}^j \right),
        \label{multiSJBD1}
    \end{equation}
where $\tilde{a}(\{\tilde{\tau}_k\})$ is a homogeneous function of degree $-1/2$ in the $\tilde{\tau}_k$ (mass dimension 2) and $\tilde{b}_{ij}(\{\tilde{\tau}_k\})$ is of degree $-5/2$ (mass dimension 10).

We now trade all dimensionful quantities for dimensionless ones as it was done in Sect.~\ref{dimsingle} and convert \eqref{multiSJBD1} to 4d Einstein frame. We can then read off the \K metric and its correction:\footnote{Note that derivatives of $K$ with respect to $\tau^i$ or $T^i$ differ only by a factor of 2. In the following, we will use the notation $\partial^2 K/(\partial\tau^i\partial\tau^j)\equiv K_{ij}$ and hence $K_{i\overline{\jmath}}=K_{ij}/4$}
\begin{equation}
		\left(S+\Delta S\right)_E = \int\dd^4x\sqrt{-g} \left[ \frac{M_{4}^2}{2}R_4 +\left( \frac{K_{ij}}{4} +  f_{ij}(\{\tau_k\})\right) \partial_\mu\tau^i\partial^\mu\tau^j \right]\,.
		\label{loopcorrgeneral}
	\end{equation}
Here $f_{ij}(\{\tau_k\})$ is derived from $\tilde{a}(\{\tilde{\tau}_k\})$ and $\tilde{b}_{ij}(\{\tilde{\tau}_k\})$ as in Sect.~\ref{dimsingle}. The functions $f_{ij}(\{\tau_k\})$ are homogeneous of degree $-4$ in the $\tau_k$. Further, $K_{ij}$ is of degree $-2$ and so \eqref{loopcorrgeneral} shows explicitly that every loop correction is necessarily suppressed by a factor of degree $-2$ in 4-cycle volumes relative to the leading term. 
Our simple dimensional analysis is in general insufficient to provide information about the dependence of $f_{ij}(\{\tau_k\})$ on individual 4-cycle volumes. However, we will be able to make progress in specific examples in Sects.~\ref{dimblowup} and \ref{fibreinflation}.

The whole argument goes through the same way using the Feynman diagram approach of Sect.~\ref{feynman}. After canonically normalizing the moduli fields, each 3-vertex (4-vertex) will again be suppressed by $1/M_4$ ($1/M_4^2$). Moreover, the argument for a possibly log-enhanced correction induced by an $R_8^4$ term on the D7-brane is still valid. The logarithmic enhancement appears in the coefficient of $R_4$ and will therefore after Weyl rescaling appear in the coefficients of all kinetic terms of the moduli. This will in turn lead to log-enhanced corrections to the \K potential.

%%%%%%%%%%%%%%%%%%%%%%%%%%%%%
\section{Warping Corrections} \label{warping}
%%%%%%%%%%%%%%%%%%%%%%%%%%%%%

In this section we discuss how warping of the
type-IIB orientifold geometry \cite{Giddings:2001yu} affects the 4d moduli action. In our 10d EFT approach below the string scale, warping corrections are simply classical backreaction effects, arising because branes and fluxes deform the CY geometry. In this sense, they are distinct from the loop corrections which are our main subject. However, concerning specifically the Kahler moduli Kahler potential, warping corrections take the form of a series of terms suppressed $1/\tau$, $1/\tau^2$ etc., where $\tau$ is a generic 4-cycle variable. This is similar to loop effects, so it is natural to include some discussion of warping in our analysis.

From a stringy perspective, the warping induced by a D-brane can be understood at leading order as a disk diagram with the boundary on the brane. More precisely, the warping far away from the brane corresponds to the regime where this disc is deformed into a long, thin cylinder, ending on the brane on one side and being capped-off by a half-sphere on the other side. Inserting, for example, two 4d graviton vertex operators in the half-sphere region gives the warping correction to the 4d Einstein-Hilbert term. An analogous discussion applies to the warping induced by an O-plane. The only difference is that the long, thin cylinder now ends in a cross-cap on one side and in a half-sphere on the other side.

The proposed association between the disk diagram and warping may at first sight appear unnatural since warping is a gravitational effect, generally associated with closed strings. However, our claim that disk diagrams on D-branes describe the leading warping effect becomes more apparent if one considers as an example a stack of D3-branes in 10d flat space in the holographic limit of \cite{Maldacena:1997re}. In the holographic limit, the open-string dynamics on the brane clearly corresponds to the closed-string or supergravity dynamics in the $AdS_5 \times S^5$ background. This $AdS_5 \times S^5$ geometry appears precisely due to the warping of the 10d flat space induced by the brane stack, consistently with our discussion above.

One way to get the first subleading order in warping is by having two disconnected disk diagrams. However, at the same order one can also have a long cylinder between two separated branes. This naturally describes the gravitational pull between two spatially separated branes. Now, since we will be interested in comparing 10d EFT loops with string loops, it is clear that the discussion of warping corrections is mandatory.

Although we will not make this concrete, one should also be able to think of the warping corrections from the perspective of Kaluza-Klein fields in a supergravity analysis. One can do so in two different ways.

In the first approach, one starts with the pure CY geometry. The KK mode expansion of 10d metric and fields is performed on the basis of this unwarped background. Introducing sources may lead to warping which, in this language, is equivalent to turning on VEVs of the 4d fields in the KK mode tower.

In the second approach, one first determines the warped geometry and performs the KK modes expansion on this basis. The resulting 4D KK tower is affected by the warping -- it is different from the tower in the first approach. The advantage is that now the VEVs of the 4d fields in the tower remain zero. One may also switch between these two perspectives by redefining the 4D KK fields. 
     
For the analysis in our paper this distinction is not relevant as we only consider loop corrections and warping separately, as independent additive effects. We expect that, for an analysis of the interplay between warping and loop effects it will be crucial to properly account for the background in which one performs the loop analysis. It would be interesting to understand such interplay effects in more detail.

The key information for us, deriving from \cite{Giddings:2001yu, Giddings:2005ff, Douglas:2008jx, Frey:2008xw, Chen:2009zi, Koerber:2007jb, Martucci:2009sf, Martucci:2014ska, Martucci:2016pzt}, is as follows: Warping induced at leading order in $\alpha'$ by fluxes, D3/O3 and curved D7/O7 branes is incorporated in the analysis of GKP \cite{Giddings:2001yu}. It corrects the Kahler potential $K$ by a series of terms $1/\tau^n$, starting at $n=1$. However, in total the tree-level no-scale structure of $K$ is not violated, such that no correction to the scalar potential arises. The statements just made follow from classical field theory. More generally, warping corrections do not represent a loop effect from our 10d EFT point of view. Nevertheless, subleading warping corrections do appear as part of a string one-loop calculation. The reader who is willing to simply accept this may move on to the next section.

The claims above may be underpinned by two series of papers which we will briefly discuss in turn.
To begin, let us write the metric as 
	\begin{equation}
		\dd s^2=e ^{2A(y,\tau)} g_{\mu\nu}(x) \dd x^\mu \dd x^\nu +e^{-2A(y,\tau)}\tilde{g}_{mn}\dd y^m\dd y^n\,,\label{wans}
	\end{equation}
where we have made it manifest that the warp factor depends on the values $\tau^i$ of Kahler moduli governing the unwarped CY geometry. However, as observed by Giddings and Maharana \cite{Giddings:2005ff}, it would be too naive to simply promote the moduli to dynamical 4d fields $\tau^i=\tau^i(x)$ since then the ansatz above does not satisfy the 10d Einstein equations. 
One has to allow for more general metric fluctuations, parametrized by so-called compensator fields \cite{Giddings:2005ff}. On this basis, the moduli space metric $K_{IJ}$ and hence the \K potential may be derived. For a deeper understanding, employing in particular the ADM/Hamiltonian formulation, see e.g.~\cite{Douglas:2008jx, Chen:2009zi}.

Using the ingredients above, the Kahler potential in the single-modulus case was derived in \cite{Frey:2008xw} (see \cite{Chen:2009zi} for a generalization allowing for mobile D3-branes):
	\begin{equation}
		K= -3 \ln\left((T+\overline{T})+2 \frac{V_W^0}{V_{CY}} \right)\,.
		\label{kwarping}
	\end{equation}
Here
\begin{equation}\label{volumeseq}
V_{CY}=\int \dd^6y\sqrt{\tilde{g}}\qquad \mbox{and} \qquad V_{W}^0=\int\dd^6y\sqrt{\tilde{g}}e^{-4A_0(y)}
\end{equation}
are the CY volume and a fiducial warped-CY volume. The real modulus $\tau=\text{Re}T$ determines the difference between general and fiducial warp factors: $e^{-4A(y,\tau)} = e^{-4A_0(y)}+\tau$. A redefinition, $T\rightarrow T'= T+V_W^0/V_{CY}$ makes it manifest that $K$ is still of no-scale form, as expected for warping corrections \cite{Giddings:2001yu}. Moreover, a large-volume expansion of \eqref{kwarping} results in a power series in $1/\tau$. Note that no factor of $g_s$ comes in since we work in the Einstein frame, such that the Poisson equation \cite{Giddings:2001yu, Giddings:2005ff} determining the warp factor $e^{-4A_0}$ contains no string coupling.\footnote{
We thank Daniel Junghans for correcting an error concerning this important point in an earlier version.
}

A generalization to the multi-moduli case has been achieved in \cite{Martucci:2014ska, Martucci:2016pzt} in a supergravity-based approach, using the earlier work \cite{Koerber:2007jb, Martucci:2009sf}. We briefly state the main ideas and results and give more details in Appendix \ref{martucci}. The first key idea of \cite{Martucci:2014ska} is to argue, on the basis of a nonlinearly realized superconformal symmetry of the 4d EFT\footnote{Extensive studies of superconformal symmetries can be found in \cite{Kallosh:2000ve}.}, that the \K potential must have the following implicit form:
	\begin{equation}
		K=K(a)= -3\ln\left( V_{CY} a \right) - 3\ln (4 \pi)\,.
		\label{K}
	\end{equation}
Here $a$ is a universal modulus which, by analogy to what has just been said in the single-modulus case, is defined as $a\equiv e^{-4A}-e^{-4A_0}$. More specifically, one may choose $A_0$ such that $V_W^0 = 0$. Now the task is to determine the functional dependence of $a$ on the chiral superfields $T^i$ conventionally used to describe the type-IIB Kahler moduli space and on possible further chiral fields, e.g.~D3-brane positions $Z_I$.

The second key idea of \cite{Martucci:2014ska} is to solve this problem by considering E3 instanton corrections: On the one hand, by holomorphicity the instanton action must be the real part of a chiral superfield. On the other hand, this action is given by the DBI action of the E3-brane in the warped background. It is determined by the warped 4-cycle volume,
\be
\frac{1}{2}\int_{D^i} e^{-4 A} J_0\wedge J_0\,,
\ee
with $J_0$ the unwarped CY Kahler form. Combining these two conditions, one arrives at
	\begin{equation}
	\begin{split}
		\text{Re}\,T^i + f^i(Z)+\overline{f}^i(\overline{Z})
		& = a \mathcal{V}^i +\frac{1}{2}\int_{D^i} e^{-4 A_0} J_0\wedge J_0 \,,
	\end{split}
		\label{chiralfields}
	\end{equation}
where $\mathcal{V}^i \equiv \frac{1}{2}\int_{D^i} J_0\wedge J_0 $ and $f^i(Z)$ are holomorphic functions of the remaining chiral fields. In the simplest case these are D3-brane positions.

From this, the desired warping-corrected Kahler potential can be derived: One first expands the  Kahler form of the unwarped Calabi-Yau as $J_0 = v^i \omega_i$, with $\omega_i$ integral harmonic $(1,1)$ forms providing a basis for $H^2(CY,\mathbb{Z})$. The $\omega_i$ are chosen to be Poincar\'e dual to the divisors $D^i$. The condition on the Kahler form 
\begin{equation}
    \frac{1}{3!} \int_{CY} J_0 \wedge J_0 \wedge J_0 = V_{CY} 
    \label{vcy}
\end{equation}
can then be thought of as a constraint on the $v^i$, which hence contain only $h^{1,1}-1$ degrees of freedom. 
Using \eqref{chiralfields} and \eqref{vcy} one can now express $a$ and the $v^i$ in terms of the variables $[\text{Re}\,T^i+f^i(Z)+\overline{f}^i(\overline{Z})]$. Inserting the resulting expression for $a$ in \eqref{K} gives the warping-corrected Kahler potential. Different choices of the constant $V_{CY}$ correspond to different additive normalizations of the $T^i$. So far, this is all rather implicit, but it suffices to make our main points. We quote a somewhat more explicit formulation in App.~\ref{martucci}. We also note that a more general calculation, including the backreaction of the \K moduli to fluxes, appears in \cite{Martucci:2016pzt}.

As demonstrated explicitly in \cite{Martucci:2014ska}, the multi-Kahler-moduli \K potential just obtained is of no-scale type. For large volumes, \eqref{K} can be expanded in $\mathcal{V}$ and the leading order correction to the \K potential is of degree $-1$ in 4-cycles. This can be seen as follows: The integral in \eqref{chiralfields} is independent of the volume modulus -- it depends only on the ratios of \K moduli. This integral is therefore suppressed by $1/ a\mathcal{V}^i(v)$ 
compared to the leading-order term $a \mathcal{V}^i$, which is of degree $1$ in 4-cycles.
With this, we have collected all the facts stated at the beginning of the present section.

%%%%%%%%%%%%%%%%%%%%%%%%%%%%%
\section{Relation to String Amplitude Calculations}\label{differentapproaches}
%%%%%%%%%%%%%%%%%%%%%%%%%%%%%

%%%%%%%%%%%%%%%%%%%%%%%%%%%%%
\subsection{String Loop Calculations and the BHP Conjecture} \label{bhkbhp}
%%%%%%%%%%%%%%%%%%%%%%%%%%%%%

In the last sections we have derived field-theoretically how loop corrections on Calabi-Yau geometries scale with the \K moduli. Let us now 
review the string loop results by Berg, Haack and K\"ors (BHK) in the torus orbifold case \cite{Berg:2005ja} and with the conjecture by Berg, Haack and Pajer (BHP) on how this might extend to CYs \cite{Berg:2007wt}. We will compare both viewpoints in Sect.~\ref{comparison}.

String loop calculations on general CYs are currently not feasible. Results are only available for torus orbifolds without flux but with, for example, D3-/D7-branes and O3-/O7-planes. Concretely, the $\mathcal{N}=2$ geometry $T^4/\mathbb{Z}_2\times T^2$ and the $\mathcal{N}=1$ geometries $T^6/(\mathbb{Z}_2\times\mathbb{Z}_2)$ and $T^6/\mathbb{Z}'_6$ were considered in \cite{Berg:2005ja}. Subsequently, BHP \cite{Berg:2007wt} conjectured how these BHK results might generalize to the CY case. Explicitly, the torus orbifold corrections and their proposed CY generalizations read
\begin{align}
	\delta K_{(g_s)}^{KK} & \,\,\sim\,\, \sum_{i=1}^{3} \frac{\mathcal{E}_i^{KK}(U,\overline{U})}{\text{Re}(S)\tau_i} & \stackrel{\text{CY}}{\longrightarrow} \qquad\quad &\delta K_{(g_s)}^{KK} \,\,\sim\,\, \sum_{a} \frac{\mathcal{C}_a^{KK}(U,\overline{U})\mathcal{T}^a(t^i)}{\text{Re}(S)\mathcal{V}}
	\label{bhpkk}
\\
	\delta K_{(g_s)}^{W} & \,\,\sim\,\, \sum_{i\neq j\neq k=1}^{3} \frac{\mathcal{E}_i^{W}(U,\overline{U})}{\tau_j\tau_k} & \stackrel{\text{CY}}{\longrightarrow} \qquad\quad &\delta K_{(g_s)}^{W} \,\,\sim\,\, \sum_{a} \frac{\mathcal{C}_a^{W}(U,\overline{U})}{\mathcal{I}^a (t^i)\,\mathcal{V}}\,.
	\label{bhpw}
\end{align}
Here $\text{Re}(S)$ is the inverse string coupling and $\tau_i$, $t^i$ are  4-cycle and 2-cycle \K moduli respectively. The functions $\mathcal{T}^a$ and $\mathcal{I}^a$ are linear in the $t^i$. The Calabi-Yau volume is $\mathcal{V}$ and $\mathcal{E}_i^{KK,W}(U,\overline{U})$, $\mathcal{C}_a^{KK,W}(U,\overline{U})$ are functions of the complex structure moduli $U$. In the Calabi-Yau case, they are unknown. The corrections are presented in the form of two different contributions, $\delta K_{(g_s)}^{KK}$ and $\delta K_{(g_s)}^{W}$, with the indices referring to `Kaluza-Klein' and `winding'. These names will be discussed in Sect.~\ref{comparison} below.

For a toroidal orbifold, the $\mathcal{I}^a$ are 2-cycles on which D7-brane stacks intersect while the $\mathcal{T}^a$ are 2-cycles transverse to the available D7-brane stacks \cite{Berg:2007wt, Cicoli:2007xp}. For a generic CY it is not obvious whether an unambiguous definition of the latter `transverse' 2-cycles exists.

In our understanding, the BHP proposal on the r.h.~side of \eqref{bhpkk}, \eqref{bhpw} consists of two steps. First, the scaling in terms of Kahler moduli is assumed not to change in going from torus orbifold to Calabi-Yau. This is rather convincing and in good agreement with the scaling arguments we discussed in previous sections, cf.~also~\cite{vonGersdorff:2005bf}. The formulae on the r.h.~side of \eqref{bhpkk}, \eqref{bhpw} would be consistent with this scaling if $\mathcal{T}^a$ and $\mathcal{I}^a$ were replaced by any homogeneous function of the 2-cycle variables of degree 1. The second part of the conjecture then states, non-trivially, that these are not just homogeneous functions but, specifically, linear expressions in the $t^i$. This linearity does not follow from our derivation in Sect.~\ref{dimmulti}, where only the homogeneity of degree 1 is obtained. In particular, extra ratios of 2-cycle volumes may appear. An example suggesting that this indeed happens is provided in Sect.~\ref{fibreinflation}. 

The loop corrections to the Kahler potential induce corrections to the scalar potential. At the perturbative level, the leading such corrections take the form \cite{Cicoli:2007xp, Cicoli:2008va}
\begin{equation}
		\delta V^{\text{1-loop}}_{(g_s)}= \left(\, \sum_{i=1}^{h^{1,1}} \frac{({\cal C}_i^{(KK)})^2}{\text{Re}(S)^2}K_{ii}^{\text{tree}}\,-\,2\delta K_{(g_s)}^W\right)\frac{W_0^2\,g_s}{\mathcal{V}^2}\,.
		\label{dVBHP}
\end{equation}
A key role in obtaining this result is played by the `extended no-scale structure' (ENSS) \cite{vonGersdorff:2005bf,Cicoli:2007xp}. This refers to the fact that the leading order contribution from corrections $\delta K$ to the \K potential vanishes if $\delta K$ is a homogeneous function of degree $-1$ in 4-cycles. Without the ENSS, one would expect a term linear in ${\cal C}_i^{(KK)}$ to be present in \eqref{dVBHP}. This term would be dominant since it would scale with the volume as ${\cal V}^{-8/3}$. Thanks to the ENSS cancellation, the Kaluza-Klein correction contributes only at second order and the leading loop correction to the potential scales as ${\cal V}^{-10/3}$.

%%%%%%%%%%%%%%%%%%%%%%%%%%%%%
\subsection{Comparing field-theoretic and (conjectured) string-theoretic Loop Effects} \label{comparison}
%%%%%%%%%%%%%%%%%%%%%%%%%%%%%
In this section we compare and match the results of our field-theoretic analysis of corrections to the \K potential (Sects.~\ref{loops} and \ref{generalization}) with the expectations from string amplitude calculations (Sect.~\ref{bhkbhp}). We will in particular suggest a resolution for a discrepancy between the field-theory analysis of \cite{vonGersdorff:2005bf} and the string amplitude results \cite{Berg:2005ja} (together with the conjecture \cite{Berg:2007wt}). This discrepancy was discussed in \cite{Cicoli:2007xp} but has, to the best of our knowledge, so far not been resolved. The discrepancy arises as follows:

From genuine loop effects we obtain corrections to the \K potential of degree $-2$ in 4-cycles. This matches the form of the BHP winding corrections. But the BHP conjecture proposes a leading correction to the \K potential, called KK correction by the authors, which is proportional to $g_s$ and of degree $-1$ in 4-cycles. Thus, it has to be clarified how this correction arises if we take the 10d EFT below the string scale as our starting point. In the remainder of this section, we argue that the EFT counterpart of the BHP KK correction are specific terms of local $\alpha'$ corrections discussed in Sect.~\ref{membranesection}. On the way, we try to develop a better physical understanding of our field theory corrections from a worldsheet perspective and vice versa.

Let us start with the interpretation of genuine loop corrections from a worldsheet perspective.
They correspond to those parts of a string 1-loop integral where the worldsheet has, roughly speaking, one long and one short dimension.
Pictorially, this means that one has a long and thin torus/Klein bottle in the closed string case or, similarly, a long and thin annulus/Moebius strip in the open string case.
Those are the regimes where the string loop integration can be identified with the field theoretic loop integral, i.e.~with the propagation of a 10d or brane-localized massless state around a loop. Comparing this interpretation with the BHP conjecture, we find that 
the two perspectives nevertheless appear to have an imperfection:

\begin{figure}[ht]
	\centering
	\includegraphics[width=0.7\textwidth]{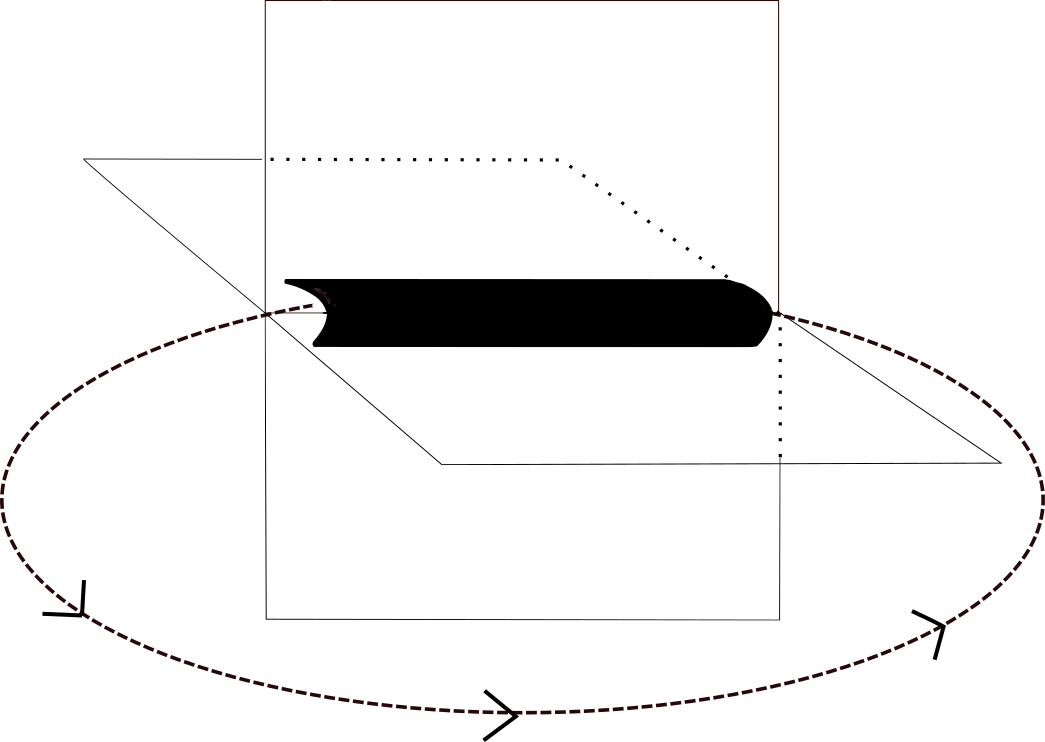}
	\caption{Short open string connecting two localized objects. The ellipse with attached arrows represents that the string is wrapped around the intersection cycle.}
	\label{winding}
\end{figure}

For the BHP winding correction to appear, it was argued in \cite{Berg:2007wt} that D7-branes or O7-planes need to intersect. Then a short open string connecting the two branes (or brane and image brane) may propagate in a closed loop along the intersection surface (see Fig.~\ref{winding}). This corresponds to the thin annulus above or, equivalently, to a field-theoretic loop effect of a massless, intersection-localized state. So far, everything looks perfect. Also the name winding correction is justified if one reinterprets the worldsheet as a closed, winding string which propagates over a short distance from brane to brane. However, as a field theorist one would expect genuine loop corrections to arise more generally -- they are not tied to intersecting objects. We have seen examples for this at the beginning of Sect.~\ref{dbranes} where an open string 1-loop effect on a single brane appears to contribute genuine loop correction. Similarly, according to Sect.~\ref{loops} closed string 1-loop effects in the bulk should also provide a loop correction of the same type and with the same scaling. It is not clear to us why the explicit string loop analysis does not see this more general type of correction
producing additional terms of degree $-2$ in 4-cycles volumes. Conceivably, this is due to the special torus based geometries underlying the calculations.

Next, we discuss local $\alpha'$ corrections. According to our definition, these are classical effects arising from the dimensional reduction of local, higher-dimension operators. However, such operators receive contributions from the high-momentum region of field theory loops. This region corresponds to string 1-loop effects where the worldsheet has a short, string-scale extension in both dimensions. There are two specific examples of this in our context 
which match the parametric scaling of the BHP KK correction: First, consider the Einstein-Hilbert term on the intersection-2-cycle of two D7-branes or of an D7/O7 pair.
This term arises from a short open string stretched from brane to brane near the intersection surface and propagating on an (also short) closed loop. Equivalently, one may think of a short, closed string exchanged between branes. The closed string carries KK momentum and one may hence call this a KK correction, as proposed in BHP. Second, we can consider the operator $M_{10}^4g_sR_8^2$ on a D7/O7. This operator can be understood as arising from a 1-loop open string diagram on the D7/O7. Equivalently, it is a short closed string emitted by the brane and absorbed by the same brane after propagating a string-scale distance.
From what has just been said, it is clear why the scaling analysis of \cite{vonGersdorff:2005bf} does not capture these local $\alpha'$ effects: While they {\it can} be interpreted as loop effect, the relevant scale is the cutoff or string scale. Thus, the in principle correct assumption that finite loop effects are dominated by the KK scale does not apply to the present contribution, which comes from the UV end of the integral.

Finally, we turn to warping effects.  As reviewed in Sect.~\ref{warping}, warping effects at subleading order scale as the BHP winding correction. Field-theoretically, warping is a classical backreaction effect and one may think of it as coming from the propagation of massless 10d fields between some source and the point where the geometry is being warped. One of the relevant string diagrams describing this is the tree-level exchange of a closed string between branes, i.e.~a long cylindrical worldsheet. Alternatively, this may be viewed as a one-loop diagram, with a long open string propagating in a short loop. Such effects should in principle be part of the analysis performed by BHK/BHP. By contrast, they are clearly not part of field-theoretic loop analyses. 
As we have discussed in Sect.~\ref{warping}, warping effects do not correct the scalar potential as they are no-scale to all orders in a large-volume expansion~\cite{Giddings:2001yu}.
For applications, it would hence be important to split the winding effect in \eqref{bhpkk} according to ${\cal C}_i^{W} = {\cal C}_{i,\,gen}^{W} + {\cal C}_{i,\,warp}^{W}$. Then the scalar potential correction of \eqref{dVBHP} would have to include only the genuine loop effect, i.e.~only ${\cal C}_{i,\,gen}^{W}$.

Let us now change perspective and check that we have identified all integration regions of a string 1-loop calculation in our field theoretic approach: An open- or closed-string 1-loop worldsheet which is short in both dimensions corresponds to a local $\alpha'$ effect. The strongest scaling is that of the BHP KK correction. A closed-string 1-loop worldsheet with one long and one short dimension corresponds to genuine loop effects, scaling like the BHP winding correction. An open-string 1-loop worldsheet corresponds either to genuine loop corrections (the case of a long strip) or to warping corrections (the case of a long cylinder). For both cases, the scaling is that of the BHP winding correction.

Finally, a worldsheet with large extension in both dimensions gives an exponentially suppressed contribution, which we can neglect and do not attempt to identify in the field theory perspective. Thus, we appear to have found all relevant regions of the integration over worldsheet geometries in our field-theoretic analysis.

Before closing, let us discuss how loops of a D7-brane gauge theory correct the Kahler modulus kinetic term. This effect has been employed in \cite{Cicoli:2007xp} to argue that genuine loop contributions exist which scale like the BHP KK correction. We will, instead, find that the analysis of this particular effect also supports our earlier conclusion that all genuine loop corrections scale like the BHP winding contribution.

\begin{figure}[ht]
	\centering
	\includegraphics[width=\textwidth]{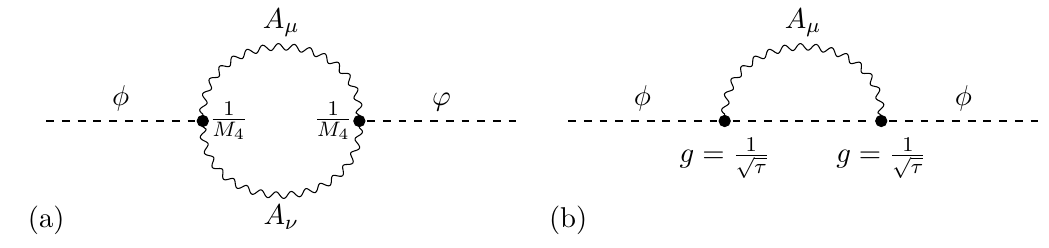}
	\caption{(a) 1-loop diagram for a D7-brane gauge field correcting the kinetic term of a modulus. (b) 1-loop diagram for the wave function renormalization of a charged scalar induced by the gauge boson.}
	\label{dbicorrection}
\end{figure}

To construct Feynman diagrams for the gauge-theory-derived loop correction to the volume modulus kinetic term, we start from gauge-kinetic term in the DBI action:
	\begin{equation}
		S_{\text{DBI}} \supset \int \dd^4x\sqrt{-g} \,\tau(x) F_{\mu\nu}F^{\mu\nu}\,.
	\end{equation}
The 4-cycle modulus $\tau$ can be expanded around its vev, $\tau(x)=\tau(1+\varphi(x)/M_4)$. Here we have chosen the fluctuation to be described by a canonically normalized scalar. Redefining $\tilde{A^\mu}=A^\mu\sqrt{\tau}$ and inserting this in the action above, we have 
	\begin{equation}
		S_{\text{DBI}} \supset \int \dd^4x\sqrt{-g}\left(  \tilde{F}_{\mu\nu}\tilde{F}^{\mu\nu}+\frac{1}{M_4}\varphi \tilde{F}_{\mu\nu}\tilde{F}^{\mu\nu}\right)\,,
		\label{dbiinteraction}
	\end{equation}
such that the 3-vertex is suppressed by $1/M_4$, as expected. Moreover, the gauge coupling is identified as $g^2=1/\tau$. 
Due to the universal suppression of the 3-vertex by $1/M_4$, we know from Sect.~\ref{loops} that the loop diagram, depicted in Fig.~\ref{dbicorrection}$\,$(a), leads to a correction which scales like BHP winding. A similar calculation has appeared earlier in the unpublished Master Thesis \cite{Roth}.

Instead, the authors of \cite{Cicoli:2007xp} estimate the loop correction by considering the wavefunction renormalization of a scalar field $\phi$ in ordinary QFT:
	\begin{equation}
	\int \dd^4x\sqrt{-g}\,\frac{1}{2}\partial_\mu \phi\partial^\mu\phi^*\qquad\to\qquad
	\int \dd^4x
	\sqrt{-g}\,\frac{1}{2}\left(1+\frac{g^2}{16\pi^2}\right)\partial_\mu\phi\,\partial^\mu\phi^*\,.
	\end{equation}
This suggests a suppression by $g^2=1/\tau$ compared to the tree level term, matching the parametric behavior of the KK corrections of BHP. 
From our point of view this has the following shortcoming:
As already noted in \cite{Cicoli:2007xp}, the analogy between modulus and charged scalar is not perfect. In the first case, the relevant 3-vertex (cf.~Fig.~\ref{dbicorrection} (a)) is $\varphi(\partial A)^2/M_4$. With an effective cutoff $M_{KK}$, this gives a correction $M_{KK}^2/M_4^2\sim 1/\tau^2$. In the second case (cf.~Fig.~\ref{dbicorrection} (b)), the 3-vertex is $g\phi^* (\partial \phi)A$, with $g\sim 1/\sqrt{\tau}$ and a log-divergent integral. While this gives a correction of order $g^2\sim 1/\tau$, it is not be applicable to our situation.
Moreover, BHP KK corrections have an additional factor $g_s$, which does not arise in the charged-scalar analogy.

%%%%%%%%%%%%%%%%%%%%%%%%%%%%%
\section{Examples and Applications} \label{applications}
%%%%%%%%%%%%%%%%%%%%%%%%%%%%%

%%%%%%%%%%%%%%%%%%%%%%%%%%%%%
\subsection{Blowup Modulus: Power Counting Result informed by Localization and Generic Volume Scaling} \label{dimblowup}
%%%%%%%%%%%%%%%%%%%%%%%%%%%%%

The LVS relies on Calabi-Yau geometries where the volume takes the form
	\begin{equation}
		\mathcal{V}= f(\tau_1,..,\tau_n)-\beta_1\tau_{s,1}^{3/2}-\cdots - \beta_m\tau_{s,m}^{3/2}\,,
	\end{equation}
with $f$ a homogeneous function of degree $3/2$ in $n$ `large' 4-cycle moduli $\tau_i$. The $m$ `small' 4-cycle moduli $\tau_{s,j}$ parametrize blowups and the $\beta_j$ are numerical constants. In what follows, we will focus on the case of a single blowup, $m=1$. But our findings generalize straightforwardly to several blowup cycles if these are sufficiently well separated in the full geometry.

\begin{figure}[htb]
	\centering
	\includegraphics[width=0.8\textwidth]{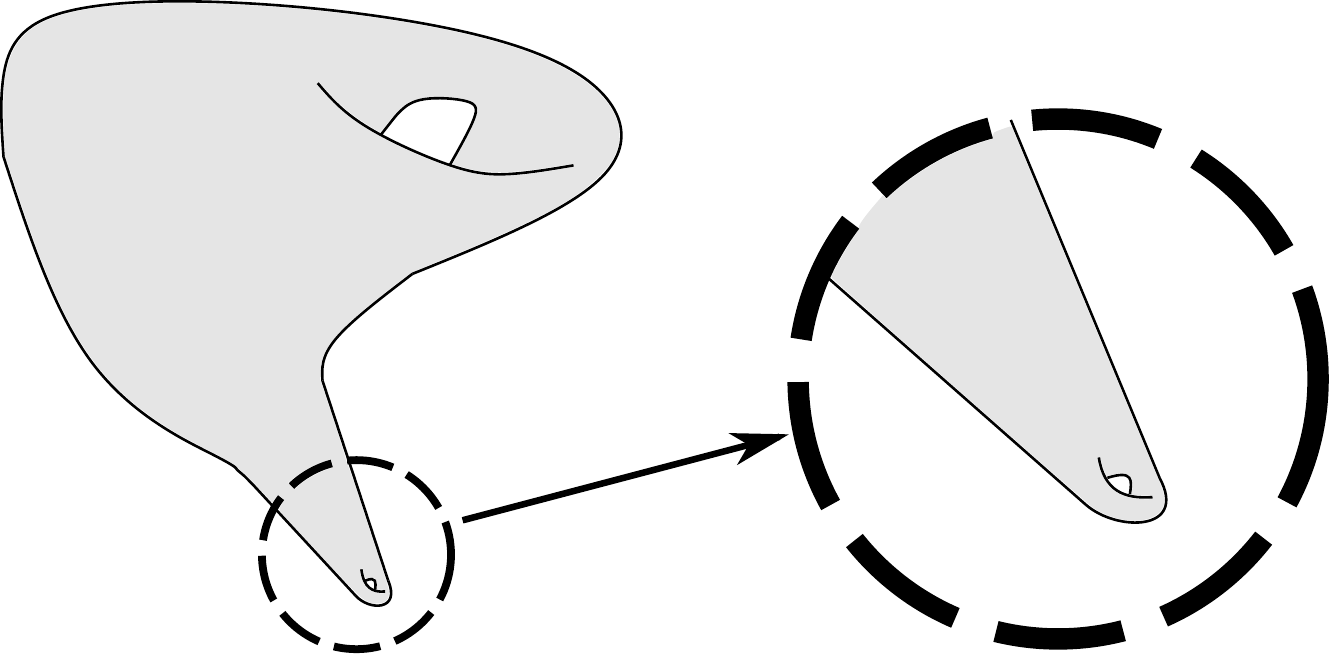}
	\caption{Illustration of a Calabi-Yau manifold with a small blowup cycle. In the vicinity of the blowup, the geometry is assumed to resemble a cone.}
	\label{blowup}
\end{figure}

The moduli stabilization mechanism of the LVS scenario ensures a hierarchical structure in the vacuum, $\tau_s\ll {\cal V}^{2/3}$. We will make the stronger assumption $\tau_s\ll \tau_i,\,\forall i$. One may then expect the geometry to be of the form illustrated in Fig.~\ref{blowup}.

Clearly, it is interesting to know the parametric dependence of loop corrections on the blowup moduli. This is important to be completely certain that loop corrections do not spoil the stabilization scenario in the first place, but it may also be useful for phenomenological applications, e.g.~to inflation \cite{Conlon:2005jm}.

In our context, a blowup is a geometric feature which induces a codimension-six singularity once the volume of the relevant 4-cycle (e.g.~a $\mathbb{C}\mathbb{P}^2$) is taken to zero. A simpler case, useful to build intuition, is the blowup of the singularity of the non-compact geometry $\mathbb{C}_2/\mathbb{Z}_2$. Famously, this is described by the explicitly known Eguchi-Hanson metric \cite{Eguchi:1978xp,Eguchi:1978gw}.
Since we are interested in 3-folds, a better model for us is the Freedman-Gibbons-Pope metric describing the blowup of $\mathbb{C}^3/\mathbb{Z}_3$ \cite{Gibbons:1979xn, Gibbons:1981}. Even closer to our case of interest is the blowup of the related compact geometry $T^6/\mathbb{Z}_3$, for which the Freedman-Gibbons-Pope metric provides an approximation.

To compute loop corrections, we assume $l_s\ll L_{\tau_s}\ll\tilde{\mathcal{V}}^{1/6}$, where $\tilde{\mathcal{V}}$ is the dimensionful Calabi-Yau volume and $L_{\tau_s}$ is the typical length scale of the $\tau_s$ cycle. An important property of the blowup modulus is that its effect on the geometry is highly localized \cite{Conlon:2011jq, Lutken:1987ny}. Specifically, in the $\mathbb{C}^3/\mathbb{Z}_3$ model the profile of the metric deformation parametrized by the blowup modulus falls off with the sixth power of the distance from the origin \cite{Lutken:1987ny}. This implies that the integral over the internal geometry which calculates the kinetic term of the modulus is of the type $\int_{\mathcal{M}_6}dy^6/(y^6)^2$ in the region $y \gg L_{\tau_s}$. Thus, the 10d dynamics of the blowup modulus is dominated by the length scale $L_{\tau_s}$. We therefore assume that it is a reasonable approximation to treat the blowup modulus as localized in the internal 6d space at a point $y_0$, which characterizes the locus of the would-be singularity.\footnote{
An alternative approach to derive loop corrections is to sum over the contributions of the KK-tower, taking into account how each mass depends on the blowup cycle. While an estimate for the lowest modes, with wavelength much larger than $L_s$ is possible \cite{Roth}, the challenge of extending such an analysis to modes with wavelength $\sim L_s$ appears daunting to us.
}

Our blowup modulus is thus identified with a localized 4d scalar field, included in the 10d action according to
	\begin{equation}
		S=\frac{1}{2\kappa_{10}^2} \int\dd^{10}x \sqrt{-g} R_{10} + \frac{1}{2}\int\dd^{10}x \sqrt{-g}\delta^{(6)}(y-y_0) \partial_\mu \phi_{s} \partial^\mu \phi_s\,.
		\label{slocal}
	\end{equation}
Here the $\delta$-function must be viewed as smeared on the scale $L_{\tau_s}$. It is easy to check that, in the compact case, the relation to the conventional 4d supergravity modulus is $\phi_s/M_4 \sim \tau_s^{3/4}/\sqrt{\cal V}$.
For simplicity, we consider the graviton as the only 10d field, but our following discussion could be repeated including further 10d degrees of freedom.
We also disregard higher-dimension operators localized at $y_0$ which can in principle induce further couplings between $\phi_s$ and the 10d metric. Their effects will not change our conclusions qualitatively.

\begin{figure}[ht]
		\centering
		\includegraphics[width=0.85\textwidth]{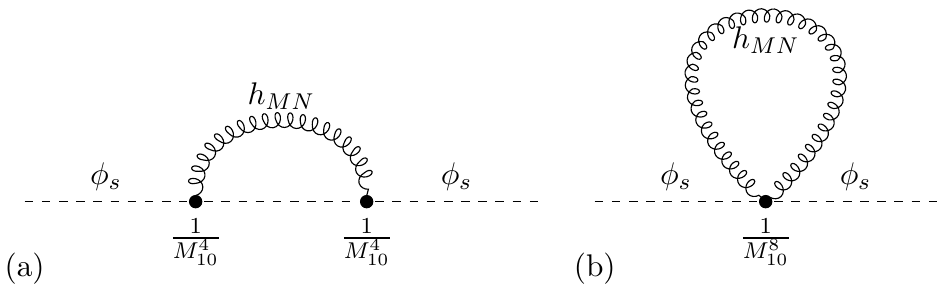}
		\caption{Self-energy (a) and tadpole (b) loop diagram correcting the kinetic term of the scalar field $\phi_s$. The scalar is confined to a 4d submanifold whereas the graviton $h_{MN}$ propagates in 10d.} 
		\label{loopsblowup}
	\end{figure}

The 1-loop diagrams correcting the $\phi_s$ propagator are shown in Fig.~\ref{loopsblowup}. To work out the corresponding integrals explicitly one has to linearize the metric, $g_{MN} = g_{MN}^{(0)} + \kappa_{10} h_{MN}$, in \eqref{slocal}. Here the background metric $g_{MN}^{(0)}$ corresponds to the singular geometry, with $\tau_s=0$. Compared to the loop analysis in Sect.~\ref{feynman}, a key difference is that the modulus is localized at a special, singular point. The propagator of $h_{MN}$ near this singularity is not known, not even approximately.
We may nevertheless make progress by using our assumption that the geometry near the singularity is conical, i.e.~in particular scale-free in the limit $\tilde{\cal V}\to \infty$. Taking this latter limit will be justified a posteriori when we see that the loops are short-distance dominated.
Moreover, we need the graviton propagator $D_h$ only with both arguments at the singularity. Thus, on dimensional grounds we have $D_h(x-x')\sim1/|x-x'|^8$, where $x,x'$ are 4d coordinates and we have suppressed any index structure. 

Using also the propagator $D_{\phi_s}\sim1/|x-x'|^2$ of the scalar $\phi_s$, one may now estimate the self-energy diagram of Fig.~\ref{loopsblowup} (a) in position space.
This diagram contributes to the correction $\delta_{\phi_s}$ to the kinetic term of the modulus, defined by ${\cal L}\supset (1+\delta_{\phi_s})(\partial \phi_s)^2/2$. Using dimensional regularization, we have 
\begin{align}
    \delta_{\phi_s}^\varepsilon&=\left.\frac{\dd}{\dd p^2}\right|_{p^2=0} \kappa_{10}^2 \mu^{-2\varepsilon}\int\dd^{4-\varepsilon}(x_1-x_2)f_{4}(\partial_{x_1},\partial_{x_2}) \text{e}^{-\text{i}(x_1-x_2)p} D_{\phi_s}(x_1-x_2)D_h(x_1-x_2)\nonumber \\
	&\sim \left.\frac{\dd}{\dd p^2}\right|_{p^2=0} \kappa_{10}^2 \mu^{-2\varepsilon}\int\dd^{4-\varepsilon}(x_1-x_2)f_{4}(\partial_{x_1},\partial_{x_2}) \frac{\text{e}^{-\text{i}(x_1-x_2)p}}{|x_1-x_2|^{2-\varepsilon}|x_1-x_2|^{8-\varepsilon}}\,,
	\label{dimregblowup}
\end{align}
where we suppressed any index structure and $f_{4}(\partial_{x_1},\partial_{x_2})$ is a homogeneous function of degree $4$ in derivative operators. The integral in \eqref{dimregblowup} has an octic UV divergence at $\varepsilon=0$.
In dimensional regularization, this integral vanishes since the only dimensionful parameter, $p^2$, is set to zero.

A second 1-loop contribution comes from the tadpole diagram Fig.~\ref{loopsblowup} (b). It is proportional to the graviton propagator in the conical geometry, $D_h$, evaluated at coincident points:
    \begin{equation}
        D_h(0)\sim\lim\limits_{x\to x'} \frac{1}{|x-x'|^8}\,.
    \end{equation}
This is a pure, octic divergence without any intrinsic mass scale, which would again give zero in dimensional regularization.

Thus, all we can learn form a dimensionally (or otherwise) regularized calculation in a low-energy EFT below the scale at which the small cycle is resolved is the following: There is no loop correction to the small-cycle kinetic term coming from the IR. Any possible corrections are dominated by the UV, i.e.~by scales at which the small-cycle geometry is resolved. Put differently, there is no interference between the deep IR, encoded in the total volume ${\cal V}$ (which we have taken to infinity when talking about a conical geometry) and the small-cycle-scale $L_{\tau_s}\sim \tau_s^{1/4}/M_{10}$.

Nevertheless, UV-dominated quantum corrections do in general exist and can simply be added by hand as a localized operator sitting at the singularity:
\be
(\partial \phi_s)^2 \,\,\to\,\,
(\partial \phi_s)^2 (1+f(\phi_s))\,.
\label{fdef}
\ee
In spite of our failure above to write down a loop integral calculating $f$ in the EFT below the scale $1/L_{\tau_s}$, we are able to determine the form of this function. We give two independent arguments: 

The first is very intuitive but it requires us to be slightly generous with the concept of a 4d EFT: Let us raise our EFT scale above $1/L_{\tau_s}$ but still below $M_{10}$. In this EFT, one starts seeing 10d supersymmetric cancellations and the loops, still described by Fig.~\ref{loopsblowup}, are cut off at the small-cycle scale: $\Lambda\sim 1/L_{\tau_s}$. The suppression by loop couplings is still governed by $\kappa_{10}^2\sim 1/M_{10}^8$. Thus,
\be
f(\phi_s)\sim \frac{\Lambda^8}{M_{10}^8}\sim \frac{1}{L_{\tau_s}^8 M_{10}^8} \sim  \frac{1}{\tau_s^2}\,, \label{fev}
\ee 
where $\tau_s \sim (\phi_s/M_{10})^{4/3}$.

The second argument is precise but slightly technical and indirect: For this, we rewrite \eqref{fdef} in terms of $\tau_s$, reinstate a finite total volume ${\cal V}$, and return to the 4d Einstein frame:
\be
(\partial \phi_s)^2(1+f) \quad\to\quad
\frac{(\partial \tau_s)^2}{\sqrt{\tau_s}}(1+f)
\quad\to\quad
\frac{(\partial \tau_s)^2}{{\cal V}\sqrt{\tau_s}}(1+f)\,.
\ee
Now we recall that, on the one hand, we have shown that genuine loop corrections to Kahler moduli kinetic terms scale with the power $-4$ in 4-cycle volumes. On the other hand, from our argument about the decoupling of IR and UV in the conical geometry near the small cycle, we know that $f$ is a function of $\tau_s$ only, independent of ${\cal V}$. This enforces $f\sim 1/\tau_s^2$, consistently with \eqref{fev}. 

As result, we have the 4d Einstein frame kinetic Lagrangian (ignoring ${\cal O}(1)$ factors)
\begin{equation}
    M_4^2\int\dd^4x \frac{1}{\mathcal{V}\sqrt{\tau_s}}\left(1+f(\phi_s(\tau_s))\right)\partial_\mu\tau_s\,\partial^\mu\tau_s\sim M_4^2\int\dd^4x \frac{1}{\mathcal{V}\sqrt{\tau_s}}\left(1+\frac{1}{\tau_s^2}\right)\partial_\mu\tau_s\,\partial^\mu\tau_s\,,
\end{equation}
Integrating the 1-loop correction twice with respect to the blowup modulus leads to a correction to the \K potential of the form
	\begin{equation}
		\delta K_{1-\text{loop}} \sim \frac{1}{\mathcal{V}\sqrt{\tau_s}}+\text{subleading terms}\,.
	\end{equation}
This matches exactly of the form of the winding corrections of the BHP conjecture. 

Note that we only expect blowup corrections of the type just discussed to arise if supersymmetry is broken to 4D $\mathcal{N}=1$ locally, near the blowup cycle. This is to be contrasted with situations where SUSY is locally $\mathcal{N}=2$ in 4D language, i.e.~all O3/O7-planes and branes are localized elsewhere in the internal geometry. Thus, the above corrections may simply not arise in LVS geometries where the blowup cycle used for moduli stabilization has locally 4D $\mathcal{N}=2$ SUSY. However, it is also possible that, if the blowup is locally 4d $\mathcal{N}=2$, nonperturbative corrections to the superpotential of the form exp($-a_s\tau_s$) are absent\footnote{Note that nonperturbative corrections to the superpotential can appear more generally than just on rigid divisors as E3-branes on effective divisors can be `rigidified' by world-volume flux and provide nonperturbative corrections \cite{Bianchi:2011qh,Bianchi:2012pn}.}, making such geometries unsuitable LVS moduli stabilization. If this were the case, then blowup cycles inducing nonperturbative superpotential effects would always come with an appropriate loop correction.

This discussion may be related to the Supersymmetric Genericity Conjecture \cite{Palti:2020qlc}. According to this conjecture, an allowed correction in a quantum gravity theory can only be systematically zero if the theory at hand descends in some way from a theory with higher supersymmetry.

We finally note that our blowup Kahler potential correction is expected to arise more generally than the similar BHP correction of `winding-type'. The latter relies on the presence of intersecting D7s. Hence, if our correction always arises in local ${\cal N}=1$ SUSY situations, and if this reduced amount of SUSY is necessary for nonperturbative effects in $W$ to appear, then this should drastically affect blowup inflation \cite{Conlon:2005jm}. The reason is that this model relies on an exponentially flat potential at $a_s\tau_s\gg 1$, which would then {\it always} be spoiled by loops, even if no D7s are present in the local geometry.

%%%%%%%%%%%%%%%%%%%%%%%%%%%%%
\subsection{Fibred Geometries and Fibre Inflation} \label{fibreinflation}
%%%%%%%%%%%%%%%%%%%%%%%%%%%%%

As we have seen in Sects.~\ref{dimmulti} and \ref{dimblowup}, determining the explicit Kahler moduli dependence of loop corrections is complicated if multiple \K moduli are present. In this section we will make partial progress in the special case of fibred Calabi-Yau manifolds, governed by two Kahler moduli. 

Fibred manifolds are of particular interest in the LVS context, where they are used for example in the Fibre Inflation proposal \cite{Cicoli:2008gp}. Fibre Inflation relies explicitly on the parametric form of loop corrections. We will see that our improved understanding of loop corrections affects Fibre Inflation.

A simple example for a fibred Calabi-Yau is provided by a K3 fibration over a $\mathbb{CP}^1$ base. Let the K3 fibre be governed by the 4-cycle modulus $\tau_f$ and the base by the 2-cycle variable $t_{b}$. To be useful in the LVS context, a blowup modulus $\tau_s$ is also needed. It will however mostly be ignored below since we have already computed the form of loop corrections associated with $\tau_s$ in Sect.~\ref{dimblowup}. The volume of this fibred Calabi-Yau takes the form 
\begin{equation}
	\mathcal{V}=t_{b}\tau_f-\tau_s^{3/2} = \frac{1}{2} \sqrt{\tau_f}\tau_2-\tau_s^{3/2}\,,
\end{equation}
where we re-expressed the base 2-cycle volume in terms of the 4-cycle modulus $\tau_2=2t_b\sqrt{\tau_f}$.
The 2-cycle volume $t_b$ (or 4-cycle volume $\tau_2$) can be traded for the total volume ${\cal V}$, as is typically done in the LVS analysis. Then the standard LVS potential stabilizes $\tau_s$ and ${\cal V}$, leaving $\tau_f$ a flat direction. In Fibre Inflation, $\tau_f$ is identified with the inflaton. Its loop-induced potential governs inflation and also stabilizes the cycle in the post-inflationary vacuum.

Building on the BHP conjecture (cf.~\eqref{bhpkk} and \eqref{bhpw}), it is argued in \cite{Cicoli:2008gp} that loops induce a Kahler potential correction
\begin{equation}
    \delta K_{(g_s)} = \frac{g_s C_1^{KK}\sqrt{\tau_f}}{\mathcal{V}}+ \frac{g_s C_2^{KK}\sqrt{\tau_2}}{\mathcal{V}} + \frac{C^W_{12}}{\mathcal{V}\sqrt{\tau_f}}\,,\label{kk2}
\end{equation}
which in turn, using \eqref{dVBHP}, induces a scalar potential for $\tau_f$ of the form
\begin{equation}
	\delta V_{(g_s)} = \left( \frac{(g_s C_1^{KK})^2}{\tau_f^2}-\frac{2C_{12}^W}{\mathcal{V}\sqrt{\tau_f}} + \frac{( g_sC_2^{KK})^2\tau_f }{2\mathcal{V}^2} \right) \frac{W_0^2\,g_s}{\mathcal{V}^2}\,.
	\label{dVfibreinflation}
\end{equation}
This assumes a geometry where D7-brane stacks wrap the 4-cycles described by $\tau_2=2 t_b\sqrt{\tau_f} $ and by $\tau_f$. The first and last term in \eqref{dVfibreinflation} are due to the KK corrections associated with the corresponding transverse cycles. The second term comes from winding corrections related to the intersection 2-cycle with volume $\sqrt{\tau_f}$. Note that the KK corrections are suppressed by $g_s^2$ compared to the winding correction. 
For such a potential, inflation corresponds to an initial situation where the fibre is much larger than the base. As the inflaton rolls, this is reversed and eventually the base is much larger than the fibre. During this process, the volume of the Calabi-Yau remains constant.

In the following, we will attempt to derive the explicit form of all loop corrections in the hierarchical regime $\tau_2\gg \tau_f \gg \tau_s$. We shall see that while all three terms with a Kahler moduli dependence as in \eqref{dVfibreinflation} will indeed appear, our result has two major differences: First, all terms in \eqref{dVfibreinflation} will arise at the same order in $g_s$. Second, our analysis suggests the presence of additional logarithmic corrections and corrections including ratios of Kahler moduli, some of which could be dominant over the usual fibre-inflation terms. Our discussion will be based on a 2-step compactification process: $10d\xrightarrow{\tau_f}6d\xrightarrow{t_b}4d$.
 
\subsubsection{Genuine Loop Effects In Fibred Geometries} \label{genuineloopfibre}

Let us first consider the 10d to 6d compactification on the fibre 4-cycle and integrate out the heavy KK modes of the fibre. 
The compact 4d manifold is governed by the single length scale $L_f\sim \tau_f^{1/4}/M_{10}$, to which we can hence apply dimensional analysis. 
We demand that both the loop correction to the Einstein-Hilbert term and the kinetic term of $L_f$,
\begin{equation}
	\Delta S_{\text{JBD}}=\int\dd^6x\sqrt{-g_6} \left[F(L_f) R_6 + G(L_f) (\partial L_f)^2\right]\,,
	\label{loopcorrfibre}
\end{equation}
are dimensionless. Since $L_f$ has mass dimension $-1$, this fixes the scaling of the prefactors to be $F(L_f)\sim 1/L_f^4$ and $G(L_f) \sim1/L_f^6$. Now we compactify to 4d, Weyl rescale by $M_{10}^8\tilde{\mathcal{V}}+L_b^2/L_f^4$ with $\tilde{\mathcal{V}}\sim L_b^2L_f^4$ to arrive at 4d Einstein frame, and express everything in terms of 4-cycle variables:
\begin{align}
    &\frac{R_6}{L_f^4} + \frac{ \left(\partial L_f\right)^2}{L_f^6} \quad 
    \xrightarrow{\text{compactify}}\quad L_b^2 \left(\frac{R_4}{L_f^4} + \frac{ \left(\partial L_f\right)^2}{L_f^6}\right) \nonumber
    \\&\xrightarrow{\text{Weyl rescale}}\quad \frac{1}{M_{10}^8}\left(\frac{1}{L_b^2L_f^4}\frac{L_b^2}{L_f^4 L_b^2} \left(\partial L_b\right)^2+\frac{1}{L_b^2L_f^4}\frac{L_b^2}{L_f^5L_b}(\partial L_b)(\partial L_f) + \frac{1}{L_b^2L_f^4}\frac{L_b^2}{L_f^6} \left(\partial L_f\right)^2 \right)\nonumber
    \\&\qquad\qquad\sim\quad \frac{\left(\partial \tau_2\right)^2}{\tau_2^2\tau_f^2} + \frac{\left(\partial \tau_f\right)\left(\partial \tau_2\right)}{\tau_2\tau_f^3}+\frac{\left(\partial \tau_f\right)^2}{\tau_f^4}\,,
    \label{10to6}
\end{align}
where we used that $M_{10}L_b=t_b^{1/2}\sim(\tau_2/\sqrt{\tau_f})^{1/2}$. We implicitly assumed that, in the compactification step from 10d to 6d, supersymmetry has been broken to 6d $\mathcal{N}=1$ by O7-planes/D7-branes wrapping the 4-cycle associated with $\tau_2=2 t_b\sqrt{\tau_f}$. We expect that the appearance of a non-trivial correction to the kinetic term is then consistent since this corresponds to 4d ${\cal N}=2$ SUSY, where corrections to the Kahler potential (or better the prepotential) are allowed.

We now explicitly translate the different terms in the last line of \eqref{10to6} in corrections to the Kahler and scalar potentials. The final term gives
\begin{equation}
\label{KahlerCorrectionFibre}
	\delta K_{(\tau_f)} \sim \frac{1}{\tau_f^2}\,.
\end{equation}
This in turn induces a correction to the scalar potential which is similar to the first term in \eqref{dVfibreinflation}, albeit without a $g_s^2$ suppression, making the correction \eqref{KahlerCorrectionFibre} more important.

The first two terms in \eqref{10to6} (if not absent due to a magical cancellation) induce a logarithmic correction in the \K potential:
\begin{equation}
    \delta K_{(\tau_f)} \sim \frac{\ln \tau_2}{\tau_f^2}\,.
    \label{dKlog10to6}
\end{equation}
We immediately recognize a problem: The correction \eqref{dKlog10to6} induces a further contribution to the kinetic Lagrangian not present in \eqref{10to6}, a term of the form $\ln(\tau_2) (\partial\tau_f)^2/\tau_f^4$. We will discuss this and related inconsistencies and how they might be resolved in Sect.~\ref{fibrelogs}.

Further genuine loop corrections can arise from D7-brane-localized fields. Concretely, if a D7-brane wraps the fibre, loop corrections to the modulus kinetic term as described by the diagram in Fig.~\ref{dbicorrection}~(a) arise. In addition, loop corrections induce a 4d Einstein-Hilbert term on the worldvolume of the brane.
The form of the corrections follows from the fact that the KK masses of the gauge fields running in the loop depend only on $\tau_f$ and not on $t_b$ (or, equivalently, the volume ${\cal V}$). One then finds
\begin{align}
    \frac{R_4}{L_f^2} + \frac{ \left(\partial L_f\right)^2}{L_f^4} \quad 
    &\to\quad \frac{1}{M_{10}^8}\left(\frac{1}{L_b^4L_f^6}\left(\partial L_b\right)^2+\frac{1}{L_b^3L_f^7}(\partial L_b)(\partial L_f) + \frac{1}{L_b^2L_f^8}\left(\partial L_f\right)^2 \right)\nonumber
    \\ &\,\sim\quad \frac{\left(\partial \tau_2\right)^2}{\tau_2^3\tau_f} + \frac{\left(\partial \tau_f\right)\left(\partial \tau_2\right)}{\tau_2^2\tau_f^2}+\frac{\left(\partial \tau_f\right)^2}{\tau_2\tau_f^3} \quad\to\quad \delta K\sim\frac{1}{\mathcal{V}\sqrt{\tau_f}} \,,
    \label{d7fibre}
\end{align}
where we also displayed the expression after Weyl rescaling and the resulting effect on the Kahler potential. The corresponding correction to the scalar potential scales like the second term in \eqref{dVfibreinflation}.

Genuine loop corrections induced by a D7 wrapping the $\tau_2$-cycle do not lead to parametrically novel corrections to $K$. As above, these corrections are obtained by gauge fields running in the loop. In particular, they induce a 6d Einstein-Hilbert term on the worldvolume of the brane of the form $R_6/L_f^4$ which hence matches the correction in \eqref{loopcorrfibre}.

A further type of loop corrections arises if one
first compactifies on the fibre to 6d and then, using the 6d classical action, considers the quantum effects of the KK modes on the base. In this case the only relevant length scale is $L_b$ and one finds
\begin{equation}
	\Delta S_{\text{JBD}}=\int\dd^4x\sqrt{-g_4} G(L_b) (\partial L_b)^2\,,
	\label{slb}
\end{equation}
with $G(L_b)\sim 1/L_b^4$ on dimensional grounds. We do not discuss the related loop corrections to the Einstein-Hilbert term since they will not induce parametrically different effects in the \K potential.

After Weyl rescaling, \eqref{slb} gives
\begin{equation}
    \frac{ \left(\partial L_b\right)^2}{L_b^4} \quad\to\quad \frac{\left(\partial L_b\right)^2}{M_{10}^8L_f^4L_b^6} \,\sim\, \frac{\left(\partial \tau_\text{2}\right)^2}{\tau_\text{2}^4}+\frac{\left(\partial \tau_f\right)\left(\partial \tau_\text{2}\right)}{\tau_f\tau_\text{2}^3}+\frac{\left(\partial \tau_f\right)^2}{\tau_f^2\tau_\text{2}^2}\,,
    \label{dLbase}
\end{equation}
where we used that $M_{10}L_b=t_b^{1/2}\sim(\tau_2/\sqrt{\tau_f})^{1/2}$. 
The first term in \eqref{dLbase} implies a corrections to the \K potential of the form
\begin{equation}
	\delta K_{(\tau_2)} \sim \frac{1}{\tau_2^2} \sim \frac{\tau_f}{\mathcal{V}^2}\,. 
	\label{dKtau2}
\end{equation}
This reproduces the \K moduli dependence of the third term in \eqref{dVfibreinflation}. Note that the correction \eqref{dKtau2} is again enhanced by a factor $1/g_s^2$ compared to expectations based on the BHP conjecture.

The last two terms in \eqref{dLbase} enforce again a logarithmic correction to the \K potential, 
\begin{equation}
    \delta K_{(\tau_2)} \sim \frac{\ln \tau_f}{\tau_2^2}\,,
    \label{dKlog}
\end{equation}
which introduces a similar problem as described below \eqref{dKlog10to6}.

\subsubsection{Local $\alpha'$ Corrections from D7-branes in Fibred Geometries} \label{fibrelogs}

Let us now study the possible effects of the $R_8^4$ operator, focusing first on the case where the D7-brane wraps the 4-cycle parametrized by $\tau_2=2 t_b \sqrt{\tau_f}$. Let us moreover assume that the coefficient of $R_8^4$ displays a non-trivial logarithmic running, as is generally expected for marginal operators. Thus, starting with an ${\cal O}(1)$ coefficient at the string scale, the coefficient grows as the energy at which we study our EFT decreases to some smaller value $\mu\gtrsim 1/L_f$. If we then compactify the D7-brane theory from 8d to 6d at the scale $1/L_f$, we find an operator $\ln(M_{10} g_s^{1/4}L_f) R_8^4$. Disregarding the small effect of the factor $g_s$ under the log, we may replace this by $\ln(\tau_f)R_8^4$. After compactification to 6d, we obtain a sum of different terms:\footnote{To avoid clutter, we keep only the logarithmic piece in the coefficient of $R_8^4$ in the following terms. In general there are additional constants, such that the reader may always replace $\ln()\,\to\, \text{const}+\ln()$.}
\begin{equation}
    \int\limits_{\mathbb{R}^{1,5}\times \sqrt{\tau_f}}\ln(\tau_f)R_8^4  \,\,\,\sim\,\,\, \sum\limits_{n=0}^{3}~ \int\limits_{\mathbb{R}^{1,5}}\ln(\tau_f)R_6^{4-n}\int\limits_{\sqrt{\tau_f}} R_2^n 
    \,\,\,\sim\,\,\, \sum\limits_{n=0}^{3}~ \int\limits_{\mathbb{R}^{1,5}}\ln(\tau_f)R_6^{4-n}L_f^{2-2n}\,.
    \label{R4dimred}
\end{equation}
Here the symbol $\sqrt{\tau_f}$ under the integral stands for the integration over the 2-cycle of the fibre wrapped by the brane. The index $n$ runs over all possibilities of how $R_8^4$ could contribute to the 6d action.\footnote{
We do not consider the case $n=4$ since we know from supersymmetry that a 6d cosmological constant will not be induced.
}
Note that we have to allow for all these possible reductions as we do not know how the indices of $R_8^4$ are contracted. Hence some (or maybe all) of the four terms could turn out to vanish when the detailed structure of the $R_8^4$ term is specified by a string amplitude calculation. 

Before compactifying further down to 4d, we have to run our 6d Lagrangian from the scale $1/L_f$ down to $1/L_b$. In this process, the $n=1$ term is special because it is a marginal operator in the 6d theory. We expect further logarithmic running, in general with a different prefactor than previously in 8d. Hence, the generic prefactor of the $R_6^3$ term just above the 4d compactification scale $1/L_b$ reads $\ln(\tau_2^\alpha\tau_f^\beta)$, where $\alpha$, $\beta$ are in principle calculable constants. For $n\neq1$ no further running occurs since the operators are relevant or irrelevant and we also do not expect subleading logarithmic divergences. We have seen this already in Sect.~\ref{feynman}. Key to this feature was the absence of massive fields in the 1-loop diagrams which also holds in the 6d theory at hand. 

After compactification to 4d, we find the following corrections to the coefficient of the Einstein-Hilbert term:
\be
\ln\tau_f  \sum\limits_{n=0,2,3} \frac{L_b^{2n}}{L_f^{2n}} \frac{L_f^2}{L_b^4} +\ln(\tau_2^\alpha \tau_f^\beta) \frac{1}{L_b^2}\,.
\label{einsteinhilbertcorrectionR4}
\ee
Focusing, as above, only on the kinetic term of $L_b$, the Weyl rescaling then produces corrections of the following type:
\bea
    &&\frac{1}{M_{10}^8}\left( \ln\tau_f  \sum\limits_{n=0,2,3} \frac{L_b^{2n}}{L_f^{2n}} \frac{L_f^2}{L_b^4} +\ln(\tau_2^\alpha \tau_f^\beta) \frac{1}{L_b^2} \right) \frac{1}{L_f^4L_b^2}\frac{(\partial L_b)^2}{L_b^2}
    \label{dLbaseR4}
    \\
    & \sim & \left(\ln\tau_f\sum\limits_{n=0,2,3} \frac{\tau_2^n}{\tau_{f}^n} + \ln(\tau_2^\alpha \tau_f^\beta) \frac{\tau_2}{\tau_f}\right)\left( \frac{\tau_f(\partial\tau_2)^2}{\tau_2^5} + \frac{(\partial\tau_2)(\partial\tau_f)}{\tau_2^4}  + \frac{(\partial\tau_f)^2}{\tau_2^3\tau_f}\right)\,.\nonumber
\eea
As an illustration, let us restrict attention to the $(\partial \tau_2)^2$ contribution and write out the sum in \eqref{dLbaseR4} explicitly for this case:
\begin{equation}
\left(  \frac{\ln(\tau_2^\alpha \tau_f^\beta)}{\tau_2^4} + \ln(\tau_f) \left(\frac{\tau_f}{\tau_{2}^5} + \frac{1}{\tau_2^3 \tau_{f}} + \frac{1}{\tau_2^2 \tau_{f}^2}\right)\right)
   (\partial\tau_2)^2\,.
    \label{dtau2}
\end{equation}

Ideally, we want to write down a Kahler potential correction from which all terms in
\eqref{dLbaseR4} and no undesired further terms follow. This can not be achieved. Let us then start our discussion by writing down a correction which induces as many of the terms in \eqref{dLbaseR4} as possible and no extra terms:
\begin{equation}
    \delta K_{(R_8^4)}\sim \frac{1+\ln\tau_f+\ln\tau_2}{\tau_2^2} + \frac{1+\ln\tau_f}{\tau_2\tau_f}+ \frac{1+\ln\tau_f}{\tau_f^2} + \frac{\tau_f(1+\ln\tau_f)}{\tau_2^3}\,.
    \label{dKR4}
\end{equation}
Here we suppressed numerical factors in each term. We may try to compare this to the structure of corrections deduced from the BHP conjecture in \cite{Cicoli:2008gp} (cf.~\eqref{kk2} above). At first sight, this is very different since (up to logs) all of our terms are homogeneous of degree $-2$ in 4-cycles. Thus they are all of `winding-type' in BHP language. Correspondingly, there are no $g_s$ factors. Interestingly, the difference becomes less dramatic at the level of the scalar potential, cf.~\eqref{dVfibreinflation}: The first three terms of \eqref{dKR4} reproduce, at the level of power-like scaling, the structure deduced from the BHP conjecture in \eqref{dVfibreinflation}. However, the additional logs could clearly be very important in concrete applications. The last term of \eqref{dKR4} contradicts the BHP conjecture at an even more elementary level in that it involves an additional 4-cycle ratio: $\tau_f/\tau_2$. 

Let us now comment on the kinetic terms for which we were not able to write a consistent Kahler potential. This concerns part of the terms in \eqref{dLbaseR4} as well as the previously emphasized problematic terms in \eqref{10to6} and \eqref{dLbase}. First, on the positive side, we note the the previously emphasized problem with the Kahler correction \eqref{dKlog} associated with \eqref{dLbase} could now potentially be resolved: The required term appears in \eqref{dKR4} and its log might thus be explained by the log associated with the $R_8^4$ operator. However, the problem with the correction \eqref{dKlog10to6} associated with \eqref{10to6} is still there: This Kahler correction induces kinetic terms that do not arise in our analysis. Moreover, problems arise for all those kinetic terms in \eqref{dLbaseR4} that lead to a product of logarithms in the appropriate correction $\delta K$. This happens for example for the last term in \eqref{dtau2}. This kinetic term would require a correction $\delta K \sim \ln(\tau_f) \ln(\tau_2) / \tau_f^2$. This in turn leads to a new kinetic term for $\tau_f$ not present in \eqref{dLbaseR4}.

We hence have to draw one of the following conclusions:

First, since in our approach it is not possible to compute numerical prefactors of each correction, the inconsistent kinetic terms could be zero due to some magical cancellation. This is not implausible since oftentimes more that a single term contributes at a specific order and the required precise compensation could then be a consequence of supersymmetry. The total loop induced correction to the \K potential including genuine loop effects is then given by \eqref{dKR4}.

Second, it could be that only the kinetic terms leading to \eqref{dKlog10to6} and \eqref{dKlog} vanish due to cancellation. The other inconsistent kinetic terms could vanish due to a particular structure of the $R_8^4$ operator and its log-divergences. The log-divergence producing the problematic kinetic terms should be absent. Assuming that all inconsistent terms produced by the log enhanced $R_8^4$ operator vanish due to the structure of the $R_8^4$ operator, only the $n=2$ case in \eqref{dLbaseR4} is left. The total correction to $K$ including genuine loop effects is then given by
\begin{equation}
    \delta K \sim \frac{1}{\tau_f^2} + \frac{1+\ln \tau_f}{\tau_2\tau_f} + \frac{1}{\tau_2^2}\,.
    \label{dKconstrained}
\end{equation}
It is of course also possible that some of the problematic kinetic terms induced by $R_8^4$ vanish due to cancellation and some by the structure of $R_8^4$. The correction $\delta K$ would then be given by 
\eqref{dKconstrained} extended by some (but not all) of the additional terms in \eqref{dKR4}.

Finally, the inconsistencies could be explained by field redefinitions which can in principle remove certain corrections to $K$. We have seen an example of this in Sect.~\ref{dbranes}. 

For a D7-brane wrapping the $\tau_f$-cycle an entirely analogous analysis of corrections coming from an $R_8^4$ term on the brane worldvolume can be carried out. This does not produce any parametrically novel effects. We finally note that, also in the fibred setting, a correction associated with the small cycle of the type $\delta K \sim 1/(\mathcal{V} \sqrt{\tau_s})$ is expected to arise. This follows in analogy to the discussion of Sect.~\ref{dimblowup}.

Let us summarize: We can reproduce the corrections to the scalar potential \eqref{dVfibreinflation} used in \cite{Cicoli:2008gp}, but with some important  differences and caveats. In our analysis, all three terms appear at the same order in $g_s$. Thus, the analogue of the hierarchy $(g_s C_1^{KK})^2, ( g_s C_2^{KK})^2/2\ll 2C_{12}^W$, which is required for fibre inflation, can be harder to realize (see~\cite{Cicoli:2016xae} for more details on the required hierarchies). Moreover, higher-curvature terms that we expect to be present on D7-branes induce log-enhanced terms of the same structure. Those would be dominant in the \K and scalar potential compared to \eqref{dVfibreinflation}. Finally, we find a term which, though small in the relevant regime $L_f\ll L_b$,
contradicts the BHP conjecture. For a deeper understanding of Fibre Inflation it is hence essential to make sure whether an $R_8^4$ term on D7-branes exists and to derive its precise structure.

Besides the corrections we have discussed so far, a further, $g_s$-suppressed effect arises if two brane stacks intersect on the 2-cycle of the fibre with volume $\sim \sqrt{\tau_f}$. It is due to the Einstein-Hilbert term induced at 1-loop order (and hence suppressed by $g_s$) on the intersection locus (see Sect.~\ref{intersectingbranes}). We may then apply \eqref{Sintersection}, generalized to the case with multiple \K moduli. The resulting correction to the \K potential, $\delta K \sim g_s\sqrt{\tau_f}/\mathcal{V}$, reproduces the first term in \eqref{kk2} and \eqref{dVfibreinflation}, including the right power of $g_s$.

\subsubsection{Loop Corrections in the Inverse Fibration} \label{inverse}

In the initial stage of Fibre Inflation, we have $\tau_f\gg\tau_2$. This situation is not directly amenable to our earlier analysis as the length scale $L_b$ of the base is smaller than the length scale $L_f$ of the fibre. One would have to first dimensionally reduce on the base to obtain an 8d theory, which could then be further compactified to 4d on the larger fibre-4-cycle. For this, one would need the geometry in this regime to possess some form of `inverse fibration' structure. In other words, one should be able to reinterpret what was originally the base 2-cycle $t_b$ as being fibred over the 4-cycle $\tau_f$. It is not clear to us whether such an inverse fibration emerges in the case at hand. Clearly, in a toy model where the 6D internal manifold is a product of a 2D and a 4D manifold, the required notion of inverse fibration trivially exists. 

Assuming the inverse fibration to exist, one can perform a similar analysis as above in the limit $\tau_f\gg\tau_2$ using $10d\xrightarrow{t_b}8d\xrightarrow{\tau_f}4d$ as the 2-step compactification process. In the following, we will only be concerned with the differences compared to the calculation above. The first difference appears in 8d, where we expect no loop corrections to the \K potential due to the larger, 8d $\mathcal{N}=1$ and hence 4d ${\cal N}=4$, supersymmetry. Next, let us include D7-branes and check whether corrections contradicting the BHP conjecture occur. This is indeed the case and it happens because extra factors involving ratios of cycles appear (cf.~the last term in \eqref{dKR4}). Concretely, such an effect is  induced by a D7-brane wrapping $\tau_2$ both through genuine loop corrections and by dimensionally reducing $R_8^4$ like 
\begin{equation}
    \int\limits_{\mathbb{R}^{1,5}\times t_b} R_8^4 \sim \int\limits_{\mathbb{R}^{1,5}} R_6 \int\limits_{t_b} R_2^3\sim \int\limits_{\mathbb{R}^{1,5}} R_6 /L_b^4 \sim \int\limits_{\mathbb{R}^{1,3}}R_4 \,L_f^2/L_b^4\,.
    \label{R4red10to6}
\end{equation}
This is analogous to the $n=0$ contribution in  \eqref{einsteinhilbertcorrectionR4}. The equivalent genuine loop effect is induced on the 6d worldvolume of the wrapped D7-brane by gauge fields running in the loop. Here we assume that the corresponding KK modes depend only on the length scale of $t_b$. This induces a localized 6d Einstein-Hilbert term on the brane worldvolume of the form $R_6 / L_b^4$, which after compactification gives the same result as on the r.h.~side of \eqref{R4red10to6}. Weyl rescaling turns this into a \K potential correction $\delta K\sim \tau_f/\tau_2^3$. This correction in now by far more interesting as it is dominant in the regime $\tau_f\gg\tau_2$ and could therefore strongly affect Fibre Inflation. We note that, similar to what has been discussed in Sect.~\ref{fibrelogs}, a log coefficient in front of \eqref{R4red10to6} would be inconsistent.

We recall that the last paragraph is to be read with the caveat that the geometry in the regime $\tau_f\gg \tau_2$ has an interpretation as an inverse fibration. This caveat disappears if we leave fibre inflation aside and simply start with a geometry which, by construction, consists of a 2-cycle with volume $t_b\sim \tau_2/\sqrt{\tau_f}$ which fibred over a base with volume $\tau_f$. The analysis of the last paragraph then applies without extra assumptions.

%%%%%%%%%%%%%%%%%%%%%%%%%%%%%
%%%%%%%%%%%%%%%%%%%%%%%%%%%%%
\section{Towards Applications in LVS and KKLT}
\label{generalissues}
%%%%%%%%%%%%%%%%%%%%%%%%%%%%%
Here, we want to collect a number of further observations concerning the role of the loop corrections we studied in concrete phenomenological scenarios. 

We first note that in the LVS, loop corrections are commonly used to stabilize non-blowup \K moduli for cases where $h^{1,1}>2$. Doing so rigorously is notoriously difficult as it requires precise knowledge of the Kahler moduli dependence of the loop correction to the scalar potential. This has so far only been achieved for torus orbifold examples \cite{Berg:2005ja} and we hope that further work along the lines of the present paper will improve this situation.

We now turn to the question of whether loop corrections are capable of upsetting the LVS. This is in principle the case and hence loop corrections have been rightfully listed as part of the corrections to be dealt with in the recent critical analysis of \cite{Junghans:2022exo}. We expect that the blowup loop correction $\delta K\sim 1/{\cal V}\sqrt{\tau_s}$, which has not been discussed in that analysis, represents one of the leading loop effects.\footnote{
In the LVS, equally important contributions to the scalar potential come from the interplay of KK-type loop corrections with nonperturbative effects, namely from $K^{s\overline{s}}\partial_s W\partial_{\overline{s}}W$, see \cite{Junghans:2022exo} chapter 3.1.1.}
It corrects the scalar potential by terms which are suppressed by $g_s^{3/2}/\sqrt{\tau_s}$ or $1/\tau_s^2$ relative to the leading LVS scalar-potential terms. For establishing an AdS minimum this is not dangerous since $g_s$ (and hence $1/\tau_s)$ can be tuned extremely small. But as has been quantified very recently in \cite{Gao:2022fdi}, after including an anti-D3-brane uplift to dS control over the LVS becomes crucially limited by the LVS Parametric Tadpole Constraint: The size of the available negative D3 tadpole limits the size of the volume and hence of the small cycle. Explicitly, one has $a_s\tau_s\lesssim 16\pi N/(9\times (12\cdots 46))$, which very significantly restricts the size of $\tau_s\sim 1/g_s$ \cite{Gao:2022fdi} and hence our ability to be safe from loop effects. Nevertheless, parametric control is clearly achievable in principle since $\tau_s$ is, by the definition of the LVS, a large parameter.

Given the previous comment, it is clearly highly relevant to determine the precise expansion parameter governing this and possibly higher loop corrections. Indeed, recall that in 4d gauge theory the true expansion parameter is $g^2/(16\pi^2)$ rather simply $g^2$. This line of thinking is known as `naive dimensional analysis' \cite{Chacko:1999hg}. In our context, the explicit study of ref.~\cite{Berg:2005ja, Berg:2005yu}\footnote{See e.g.~eqs.~(10), (12) of \cite{Berg:2005yu} and eq.~(2.6) of \cite{Broy:2015zba}.} suggests that the expansion parameter is 
$1/(2\pi)^4\tau^2$, with $\tau$ being a generic 4-cycle variable. (Relating this to the previous paragraph, we would have $\tau^2\sim {\cal V}\sqrt{\tau_s}$.) To the best of our present understanding, part of this significant suppression by $(2\pi)$ factors is associated, in field theory language, with the explicit results for sums over KK modes on tori (see e.g.~\cite{Cheng:2002iz}, Sect.~II). It would be interesting to understand whether the $1/(2\pi)^4$ factor survives in CY geometries. This is not obvious since the loop factor in a purely 4d approach (following the logic of Sect.~\ref{feynman}) is $\sim \Lambda^2/(16\pi^2 M_4^2)$. Identifying the cutoff with the KK mass scale, one does of course get the expected parametric behaviour $m_{KK}^2/M_4^2\sim 1/\tau^2$, but fixing the $(2\pi)$ factors in this relation appears difficult in a CY geometry.

We have seen that small 4-cycles $\tau_s$ induce loop corrections to the Kahler metric which are $1/\tau_s^2$ suppressed compared to the leading terms, see Sect.~\ref{dimblowup}. Clearly, such corrections are dangerous when the small-cycle volume becomes $\mathcal{O}(1)$ in Einstein frame. More generally, going beyond the specific case of LVS-type blowup-cycles, one might suspect that even corrections suppressed only by 2-cycle volumes exist. For example, the previously discussed blowup correction came from a term $\delta K\sim 1/{\cal V}\sqrt{\tau_s}$, where the 2-cycle volume $\sqrt{\tau_s}$ appears. Now, in a series of papers \cite{Demirtas:2019sip, Demirtas:2020ffz, Demirtas:2021nlu, Demirtas:2021ote}, flux compactifications with large $h^{1,1}$ were constructed with the goal to obtain explicit KKLT-type models. While, as already noted in \cite{Gao:2020xqh}, having large $h^{1,1}$ appears to be a promising route to counteract the `singular-bulk problem', this may force one into a regime of dangerously small 2-cycles. Indeed, in \cite{Demirtas:2021nlu} many 2-cycles with Einstein-frame volumes of order unity or smaller arise. While this may be harmless in cases where, as the authors argue, the shrinking of the 2-cycle merely leads to a conifold singularity with local ${\cal N}=2$ SUSY, it is not clear whether the presence of nearby O-planes with reduced SUSY can always be avoided. If it can not, then our loop corrections with potentially only a 2-cycle Einstein-frame-volume suppression represent a serious concern.

Moreover, the above compactifications have many, typically $\mathcal{O}(100)$, small cycles. One might then be concerned that even if individual small-cycle corrections are controlled, the corrections could add up to become dangerously large if many small 2- or 4-cycles contribute. It would be interesting to better understand the form these small-cycle loop corrections in settings where they are not well-separated.

Let us now change perspective and estimate from a 4d EFT perspective and without the detour via the Kahler metric at which order in the inverse volume $1/\mathcal{V}$ loop corrections to the scalar potential arise. Due to the supersymmetric spectrum, i.e.~$\text{Str} \mathcal{M}^0=0$, the quartic divergence $\sim \Lambda^4$ vanishes and we have
(disregarding numerical prefactors)
\begin{equation}
    V_{1-\text{loop}} \sim V_\text{tree}
    + \Lambda^2\,\text{Str} \mathcal{M}^2+  \text{Str} \mathcal{M}^4\, \ln\left(\frac{\Lambda^2}{\mathcal{M}^2}\right)+\dots\,.
    \label{colemanweinberg}
\end{equation}
Here $\Lambda$ is the cutoff and $\mathcal{M}^2$ the mass matrix. We also assume that SUSY is broken in a flat or approximately flat background (as e.g.~in the LVS). This may be quantified by requiring $1/L_{\rm AdS/dS}\ll m_{3/2}$ and it implies that the supertrace of the mass matrix obeys $\text{Str} \mathcal{M}^2 \sim m_{3/2}^2$ (see e.g.~\cite{Grisaru:1982sr, Ferrara:1994kg}). In our case, $\Lambda=m_{KK}\sim M_4/\mathcal{V}^{2/3}$ and $m_{3/2}\sim M_4 g_s^{1/2} W_0/\mathcal{V}$. For sufficiently large volume, the SUSY breaking scale is hence below the KK scale, as required for consistency of our 4d analysis.
Using this, the leading correction is given by
\begin{equation}
    V_{1-\text{loop}} 
    \sim  V_\text{tree} + M_4^4\frac{g_sW_0^2}{\mathcal{V}^{10/3}}+\dots\,.
    \label{1ld}
\end{equation}
This precisely matches the order at which the genuine loop effects correct the scalar potential (cf.~Table \ref{tab:correction_summary}). Independently, this represents a very general argument suggesting that the leading loop effect in the scalar potential scales as $\sim {\cal V}^{-10/3}$. This appears to clash with the existence of loop corrections $\sim {\cal V}^{-8/3}$ which have recently been proposed in the literature \cite{Burgess:2022nbx}. Note also that the Kahler potential term inducing such corrections is $\delta K\sim \ln(\tau)/\tau$, with $\tau \sim {\cal V}^{2/3}$. Consistently with what was said before, this does not correspond to a genuine loop but rather to a local $\alpha'$ correction. The log would then be expected if the underlying operator is marginal in 6d, 8d or 10d. However, we have seen that marginal operators always produce Kahler potential corrections which scale as $1/\tau^2$, not $1/\tau$. So we are left with a contradiction.

A possible resolution is as follows: Let us start with the $G_4^2 R_{11}^3$ term in 11d which according to \cite{Weissenbacher:2019mef, Klaewer:2020lfg, Cicoli:2021rub} is the origin of the effect. Let us assume that, using the curvature of the torus fibration of F-theory, this descends to a local operator $\sim (\nabla H_3)^2$ on D7-brane stacks (see e.g.~\cite{Garousi:2009dj}).\footnote{
Another option would be terms of the type $R_8 (\partial F_2)^2$ computed in \cite{Jalali:2015xca}. Note that $G_4$ flux in M-theory descends to $H_3$, $F_3$ and $F_2$ flux in type IIB. 
} 
This would lead to a term $\sim |W_0|^2$ in the 4d scalar potential. The volume scaling follows by noting that $H_3\sim 1/\tau^{3/4}$, $\nabla \sim 1/\tau^{1/4}$, $\int_{D7} d^4y \sim \tau$ and that, finally, the Weyl rescaling to the 4d Einstein frame gives a factor $1/{\cal V}^2$. In total, one finds $|W_0|^2/{\cal V}^{8/3}$.

To summarize, the proposal is that the correction $\sim \ln(\tau)/\tau$ in the Kahler potential comes from a field redefinition and has nothing to do with a marginal operator. It does, however, produce a manifestly physical correction to the scalar potential $\sim |W_0|^2/{\cal V}^{8/3}$ which, as we saw earlier, can not be understood as a 4d loop effect. Instead, it comes from a local $\alpha'$ correction, for example a brane-localized 4-derivative term involving flux and hence proportional to $|W_0|^2$.
Clearly, this is at the moment only a suggestion and it deserves further study how a possible $\ln(\tau)/\tau$ term and the corresponding scalar-potential effect are to be understood in a 10d SUGRA analysis of type-IIB orientifolds.

We leave it as a challenge for the future to study possible corrections associated with a field redefinition $\tau'_s \equiv \tau_s + \alpha \ln (\mathcal{V})$ \cite{Conlon:2010ji} in our approach (see also the comment in \cite{Gao:2022fdi}).

%%%%%%%%%%%%%%%%%%%%%%%%%%%%%
%%%%%%%%%%%%%%%%%%%%%%%%%%%%%
\section{Discussion} \label{discussion}
%%%%%%%%%%%%%%%%%%%%%%%%%%%%%
%%%%%%%%%%%%%%%%%%%%%%%%%%%%%

In this work, we have analysed corrections to the Kahler potential in type-IIB string compactifications on Calabi-Yau orientifolds. We relied on one-loop field theory together with the available information about higher-mass-dimension local operators, both in 10d and on branes or brane intersections.
The key corrections are summarized in Table~\ref{tab:correction_summary}.

A novel proposal suggested by our analysis is that log-enhanced corrections, dominant w.r.t.~established terms, arise on the basis of marginal higher-curvature operators in 8d or in 6d. The operators in question are localized on D7-branes/O7-planes or on their intersection loci respectively. One example is an $R_8^4$ operator, which is expected to be present on D7/O7s~\cite{Bachas:1999um}, another is the $R_6^3$ operators on the 6d intersection loci.
The correction affects the scalar potential at order $\ln(M_{10}g_s^{1/4}L)\times h_{-5}$, where $h_{-5}$ is a homogeneous function of degree -5 in 4-cycle \K moduli and $L$ is a typical length scale of the relevant cycle. Because of its log-enhancement, this correction tends to be dominant and may hence be critical in moduli stabilization schemes relying on loop corrections, such as in Fibre Inflation (cf.~Sect.~\ref{fibreinflation}) or, more generally, in multi-moduli LVS constructions. Moreover, log-enhanced corrections could be dangerous for the LVS per se if an uplift to dS is included. It would hence be essential to confirm the existence of $R_8^4$/$R_6^3$ terms localized on 8d/6d from a string amplitude calculation.

One of our declared goals was to derive the Berg-Haack-Pajer conjecture, which we have partially achieved: First, we understand that `winding-type' terms, correcting the Kahler potential at order -2 in 4-cycle variables, come from genuine loop effects and may, as just noted, feature a UV-sensitive log-enhancement associated with local $\alpha'$ corrections. Additional terms with the same volume dependence but suppressed by $g_s$ are expected. We were not able to confirm the more specific suggested form $\sum_a 1/\mathcal{I}_a(t^i){\cal V}$ with $\mathcal{I}_a(t^i)$ linear combinations of 2-cycle variables $t^i$. On the contrary, fibred examples suggest that a more general Kahler moduli dependence is possible.

Second, we understand the `KK-type' terms as coming from higher-curvature operators. This resolves a discrepancy between the field theory analysis of 1-loop corrections to the \K potential in \cite{vonGersdorff:2005bf} on the one hand and string loop calculations by BHK \cite{Berg:2005ja} and the BHP conjecture \cite{Berg:2007wt} on the other hand.
The local operators, like for example an Einstein-Hilbert term induced at 1-loop level on intersection cycles of D7-branes, induce KK-type Kahler potential corrections which, in turn, may modify the scalar potential.

Returning to the corrections of winding-type, we note that our analysis suggests a different interpretation than what is usually found in the literature: 
First, part of them is due to warping and, because warping respects the no-scale structure, this part will not affect the scalar potential. Second, the remaining part of the winding-type corrections corresponds to genuine loop effects and 
should hence be present more generally than proposed by BHP. The reason is that these effects are not tied to intersecting D7-branes but only to the requirement that the relevant tower of KK modes running in the loop displays an ${\cal N}=1$ rather than an ${\cal N}=2$ SUSY spectrum.
It is conceivable that the absence of corresponding contributions in \cite{Berg:2005ja} is due to the special torus geometries used. We also note that winding-type corrections (and in particular genuine loop effects) contribute to the scalar potential with a $1/g_s^2$ enhancement compared to KK-type corrections since the latter are subject to the extended no-scale structure \cite{vonGersdorff:2005bf, Berg:2007wt, Cicoli:2007xp}. Thus, genuine loop effects should always be included in scenarios where KK-type corrections play a role, like in fibre inflation \cite{Cicoli:2008gp} or $\alpha'$ inflation \cite{Broy:2015zba}. Another example where this proposed more general occurrence of genuine loop corrections is important is blowup inflation \cite{Conlon:2005jm}: For this scenario, the loop correction to a blowup cycle $\tau_s$ calculated in Sect.~\ref{dimblowup} is dangerous and, in our understanding, it is expected to always be present as long as the geometry is $\mathcal{N}=1$ locally, near the blowup cycle.
We hope that it will be possible to clarify this further, strengthening the proposal of blowup inflation or ruling it out.

Let us close with an optimistic outlook. Naively, one might fear that loop effects will never be explicitly calculable on Calabi-Yaus and will hence always be in the way of fully controlled models. But things could be much better: It is conceivable that, as we argued in this paper, the dominant loop effects in the scalar potential will always come from log-enhanced  winding-type corrections. Those are UV sensitive, being tied to certain marginal local operators. If the coefficients of the latter can be determined and the integrals over these operators in the classical Calabi-Yau background can be calculated, one may hope that the dominant effect from loops on moduli stabilization will become accessible.

\section*{Acknowledgements}

We would like to thank Michael Haack, Daniel Junghans, and Fernando Quevedo for helpful discussions. We thank an anonymous referee for valuable comments on an earlier draft of this paper.

X.G. was supported in part by the Humboldt Research Fellowship and NSFC under grant numbers 12005150. This work was supported by the Graduiertenkolleg ‘Particle physics beyond the Standard Model’ (GRK 1940) and the Deutsche
Forschungsgemeinschaft (DFG, German Research Foundation) under Germany’s Excellence Strategy EXC 2181/1 - 390900948 (the Heidelberg STRUCTURES Excellence Cluster). A.H. acknowledges many useful insights from a collaboration with Ingo Roth in the context of his Master research project \cite{Roth}.

\appendix

%%%%%%%%%%%%%%%%%%%%%%%%%%%%%
%%%%%%%%%%%%%%%%%%%%%%%%%%%%%
\section{The Warped \K Potential in the Multi-Moduli Case} \label{martucci}
%%%%%%%%%%%%%%%%%%%%%%%%%%%%%
%%%%%%%%%%%%%%%%%%%%%%%%%%%%%
In Sect.~\ref{warping} we reviewed some of the results of \cite{Martucci:2014ska}. A key starting point was the simple expression \eqref{K} for the Kahler potential in terms of the universal modulus $a$. Through the identification
	\begin{equation}
	\begin{split} \label{appendixstart}
		\text{Re}\,T^i + f^i(Z)+\overline{f}^i(\overline{Z})
		& = a \mathcal{V}^i +\frac{1}{2}\int_{D^i} e^{-4 A_0} J_0\wedge J_0 \,,
	\end{split}
	\end{equation}
it is then possible to express $a$ as a function of the other Kahler moduli. In this appendix we add some details concerning how the integral in \eqref{appendixstart} can be evaluated. This material is entirely a review of \cite{Martucci:2014ska}, to which we refer the reader for further information.

First, one observes that the harmonic basis 2-forms may be expressed through so-called $h^{1,1}$ local potentials $\kappa^i (z,\overline{z};v)$:
\begin{equation}
    \omega^i = i \partial \overline{\partial} \kappa^i (z,\overline{z};v)\,.
\end{equation}
The $\kappa^i (z,\overline{z};v)$ cannot be defined globally as the $\omega^i$ are, by definition, nontrivial in cohomology. This can be remedied by introducing sections $\zeta^i$ of the line bundles $\mathcal{O}_{CY}(D^i)$, such that $\zeta^i=0$ identifies the location of $D^i$. It can then be shown that the combination
\begin{equation}
    \pi \kappa^i - \text{Re} \log \zeta^i
\end{equation}
is globally well-defined.

With this, one can derive that
\begin{equation}
    \frac{1}{2}\int_{D^i} e^{-4 A_0} J_0\wedge J_0 = \frac{1}{2\pi l_s^4} \int_\text{CY} \left( \pi \kappa^i - \text{Re} \log \zeta^i \right) Q_6\,.
    \label{integral}
\end{equation}
Here $Q_6$ denotes the D3 charge distribution, which by itself would of course integrate to zero on the compact space. One may isolate from it the mobile D3-brane contribution by writing
\begin{equation}
    Q_6 = l_s^4 \sum\limits_{I \,\in\, \{\text{D3s}\}} \delta_I^{(6)} + Q_6^\text{bg}\,,
    \label{q6}
\end{equation}
where $I$ labels the mobile D3-branes. The background D3 charge distribution 
$Q^{\text{bg}}_6 $ includes the contributions from bulk fluxes and O3-planes:
\begin{equation}
    Q^{\text{bg}}_6 \equiv F_3 \wedge H_3 - l_s^4 \sum_{J \,\in\, \{\text{O3s}\}} \frac{1}{4} \delta_J^{(6)}\,.
\end{equation}
We could include here the D3 charge of curved D7/O7s but will not do so for notational simplicity. Inserting $Q_6$ in \eqref{integral} and introducing the functions
\begin{equation}
    h^i \equiv \frac{1}{2\pi l_s^4} \int_{CY} \left( \pi \kappa^i - \text{Re} \log \zeta^i \right) Q^{\text{bg}}_6
\end{equation}
then leads to the final expression
\begin{equation}
\text{Re} T^i = a\mathcal{V}^i(v)+h^i(v)+\frac{1}{2}\sum_{I \,\in\, \{\text{D3s}\}}\, \kappa^i(Z_I,\overline{Z}_I;v)\,.
\label{final}
\end{equation}
The last term arises since the $\delta$-function part of \eqref{q6} evaluates $\kappa^i (z,\overline{z};v)$ at the positions $Z_I$, $\overline{Z}_I$ of the mobile D3-branes. The $\text{Re} \log \zeta^i(Z_I)+ \text{Re} \log \overline{\zeta}^i(\overline{Z}_I)$ term coming from the mobile D3-branes is of the form $f^i(Z)+\overline{f}^i(\overline{Z})$ and is absorbed in the definition of these functions in \eqref{appendixstart}, hence this term does not appear in \eqref{final}.
The $h^i$ then encode the data about the geometry, bulk fluxes, and all localized objects except mobile D3-branes.

One should view \eqref{final} as a system of $h^{1,1}$ equations defining $a$ and $v^i$ in terms of $\text{Re} T^i$, $Z_I$, and $\overline{Z}_I$. In \cite{Martucci:2014ska}, the expression for $a$ in terms of $\text{Re} T^i$, $Z_I$, and $\overline{Z}_I$ is explicitly worked out in several examples. For us, the importance of the result \eqref{final} is that it may, in principle, be used to study the parametric dependence of the warping correction to $K$ on the $T^i$. One would need to understand in more detail how the last two terms on the r.h.~side of \eqref{final}, which are of degree zero in the Kahler moduli, depend on ratios Re$T^i/$Re$T^j$. 
This may then be used to disentangle which part of the winding correction comes from warping and which from genuine loop effects. Moreover, a non-trivial check of the BHP conjecture about the form of the winding correction may become possible. We leave these open problems for future work.

%%%%%%%%%%%%%%%%%%%%%%%%%%%%%
%%%%%%%%%%%%%%%%%%%%%%%%%%%%%
\bibliographystyle{JHEP}
\bibliography{refs}
%%%%%%%%%%%%%%%%%%%%%%%%%%%%%
%%%%%%%%%%%%%%%%%%%%%%%%%%%%%

%%%%%%%%%%%%%%%%%%%%%%%%%%%%%
%%%%%%%%%%%%%%%%%%%%%%%%%%%%%
\end{document}